\documentclass[aps,prb,twocolumn]{revtex4-1} %preprint
\usepackage{graphicx}
\usepackage{hyperref}
\usepackage{amsmath}
\usepackage{amssymb}
\usepackage{mathrsfs}

\providecommand{\abs}[1]{\lvert#1\rvert}

\newcommand{\ket}[1]{\left\lvert #1 \right\rangle}

\begin{document}

%\preprint{}

\title{Atomic delocalisation as a microscopic origin of two-level defects in Josephson junctions}



\author{Timothy C. DuBois}
\email[]{timothy.dubois@rmit.edu.au}
\affiliation{Chemical and Quantum Physics, School of Applied Sciences, RMIT University, Melbourne, 3001, Australia}

\author{Salvy P. Russo}
\email[]{salvy.russo@rmit.edu.au}
\affiliation{Chemical and Quantum Physics, School of Applied Sciences, RMIT University, Melbourne, 3001, Australia}

\author{Jared H. Cole}
\email[]{jared.cole@rmit.edu.au}
\affiliation{Chemical and Quantum Physics, School of Applied Sciences, RMIT University, Melbourne, 3001, Australia}


\date{\today}

\begin{abstract}
Identifying the microscopic origins of decoherence sources prevalent in Josephson junction based circuits is central to their use as functional quantum devices. Focussing on so called ``strongly coupled'' two-level defects, we construct a theoretical model using the atomic position of the oxygen which is spatially delocalised in the oxide forming the Josephson junction barrier. Using this model, we investigate which atomic configurations give rise to two-level behaviour of the type seen in experiments.  We compute experimentally observable parameters for phase qubits and examine defect response under the effects of applied electric field and strain.
\end{abstract}

%\pacs{}
%\keywords{}

\maketitle

\section{Introduction and Concept}
Superconducting qubits and Josephson junction based quantum devices in general are often limited by decoherence sources. A common and important source of decoherence is the environmental two-level system (TLS)~\cite{Dutta1981, Shnirman2005}. Experiments have probed these defects directly and shown them to be stable, controllable and have relatively long decoherence times,~\cite{Simmonds2004, Neeley2008, Shalibo2010, Lupascu2009, Lisenfeld2010} although little is known about their true microscopic nature. Many phenomenological theories attempting to describe them exist; including charge dipoles~\cite{Martinis2005}, Andreev bound states~\cite{deSousa2009}, magnetic dipoles~\cite{Sendelbach2008}, Kondo impurities~\cite{Faoro2007} and TLS state dependence of the Josephson junction (JJ) transparency.~\cite{Ku2005}Detailed fitting of experimental data can place limits on these models~\cite{Cole2010} but the scope of free parameters of each model allows them all to fit experimental results - rendering them presently indistinguishable. It is therefore important to construct microscopic models of these systems to increase our understanding of their composition. Polaron dressed electrons~\cite{Agarwal2013} and surface aluminium ions paramagnetically coupling to ambient molecules~\cite{Lee2014} are two recent models that may shed new light in this area.

Bistable defects in glasses and amorphous solids in general have been understood for some time~\cite{Anderson1972}. Amorphous insulating barriers (either in the form of Josephson junctions or simply a native oxide) form an integral part of superconducting circuits, so it comes as no surprise that TLSs are often considered to be an important source of noise in these circuits~\cite{Dutta1981, Shnirman2005, Martinis2005}. Developments in controllable qubit architecture (charge, flux and phase) has enabled the study of so-called `strongly coupled defects'~\cite{Neeley2008, Lupascu2009, Lisenfeld2010}. These defects have comparable resonance frequencies to the qubit circuit and coupling strengths as well as decoherence times long enough to allow coherent oscillations between the qubit and TLS. Probing individual defects has promoted their bistable nature from hypothesis to observable fact, as well as providing clues to their microscopic origin.

As described in previous work~\cite{DuBois2013}, we consider the origin of some defects to be within the amorphous oxide layer itself, specifically an oxygen atom in a spatially delocalised state. This has important ramifications for materials science based efforts to reduce the effects of TLSs. If, as alternative models suggest, TLS defects indeed stem from surface states~\cite{Choi2009} or the accidental inclusion of an alien species~\cite{Jameson2011, Holder2013}; future fabrication techniques or more robust qubit designs may suppress or diminish the response of such noise sources as has been the case historically~\cite{Vion2002, Martinis2005, Koch2007, Schreier2008, Houck2008}. Nevertheless, if the oxygen atoms themselves form a noise source, a perfectly clean but amorphous dielectric may still harbour a large ensemble of TLSs.

\begin{figure}
  \centering
  \includegraphics[width=\columnwidth]{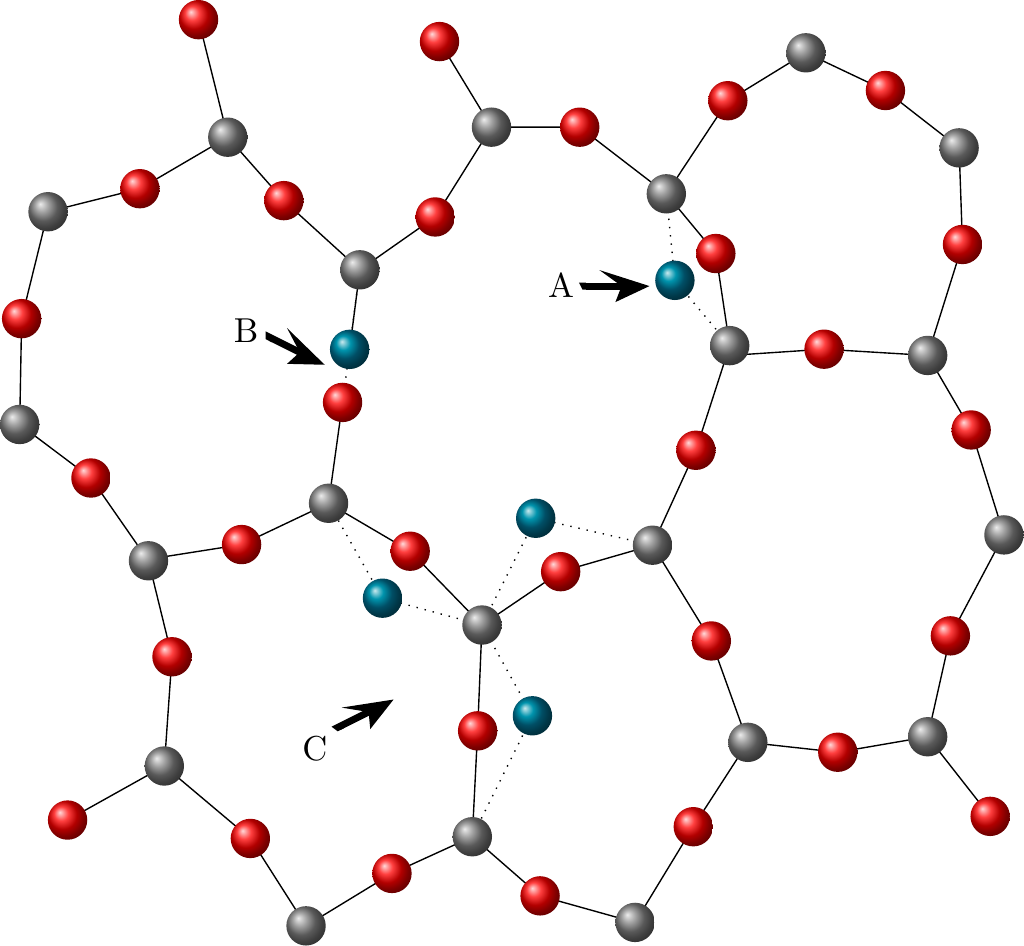}\\
  \caption{\label{fig:defects}(Color online) Illustration of a trigonal-like oxide lattice (aluminium in gray, oxygen in red) with three possible defect types~\cite{Zachariasen1932, Enss2005}. A: the bond length is shortened causing an oxygen to form a dipole perpendicular to the bond axis, B: the bond length is lengthened causing an oxygen to form a dipole parallel to the bond axis, C: a cluster of three oxygens are rotated about a central metal atom. Types A and B are of interest to this work. Type C requires a many body investigation that is beyond the current scope, although it has been discussed extensively in the literature~\cite{Buchenau1984,Heuer1998,Trachenko2000,Reinisch2005}.}
\end{figure}

Our approach considers only atomic positions as input parameters in an attempt to construct a microscopic model rather than a phenomenological one. The premise of our TLS model is that positional anharmonicity of oxygen atoms arises within the AlO$_{x}$ barrier of the Josephson junction primarily due to its non-crystalline nature. As an illustrative example, consider an interstitial oxygen defect in crystalline silicon: the harmonic approximation for atomic positions cannot be applied due to the rotational symmetry of the defect as oxygen delocalises around the Si-Si bond axis~\cite{Artacho1995}. This forms an anharmonic system with a quasi-degenerate~\cite{DuBois2013} ground state, even in a ``perfect'' crystal. This ansatz allows the existence of many different spatial configurations throughout the layer, causing unique TLS properties based solely on atomic positions and rotation in relation to the external electric field. To simplify the configuration space we initially consider the introduction of small lattice irregularities from an idealised trigonal-like 2D oxide lattice. Possible defects of this nature are depicted in Figure \ref{fig:defects}.

%Finally, recent simulations suggest a TLS mass of order $1-100 \: amu$~\cite{Nugroho2013a}, too large to be simply due to an electron moving between states. \todo[color=blue!40,inline]{This paper is out now, so reference from march meeting has been updated. WKB calculations don't seem to be there. Should we remove this?}

It has often been suggested that a bath of TLSs are responsible for the weakly coupled, ohmic $1/f$ noise~\cite{Dutta1981}. However, it is unclear if the identified strongly coupled systems are from the same origin. Ultimately many separate microscopic suspects may be identified; although work in this area is not mature enough to speculate. The model in this work therefore only attempts to describe TLSs that are strongly coupled in nature.

The outline of this paper is as follows. Section \ref{sec:methods} introduces the theory from which we build our model to investigate delocalised oxygen based two-level systems. The following three sections then start from a minimal example of this model and slowly add complexity, so interactions can be examined and understood in a systematic way. Section \ref{sec:bonds} considers a defect comprising of an Al--O--Al chain in one dimension, perturbed from a crystalline lattice, simulating defects A and B in Figure \ref{fig:defects}. In reality, an oxygen atom in the amorphous layer of a Josephson junction will be surrounded by atoms in all three dimensions. Moving towards a model representation of this, Section \ref{sec:2d} extends the model to two dimensions, with four aluminium atoms confining an oxygen in a plane. The completed model is then described in Section \ref{sec:tls}, which extends oxygen confinement into three dimensions (with six aluminum atoms), whilst still projecting oxygen delocalisation on a plane. Although in general an oxygen can delocalise in three dimensions, for this investigation we focus on an effective $2\!+\!1$D model, minimising both computational and descriptive complexity while still modelling the relevant behaviour of the system. The following sections then apply the $2\!+\!1$D model and compare delocalised oxygen responses to experimental TLS data. Section \ref{sec:smax} discuses qubit--TLS coupling and Section \ref{sec:strain} observes the effect of mechanical strain.

\section{Methods}\label{sec:methods}

We are interested in what happens to the oxygen atom as it responds to an external potential exerted via its' nearest neighbours, particularly if the bonded aluminium atoms are displaced in a manner similar to the defect types outlined in Figure \ref{fig:defects}. If we also ignore any time evolution properties of the system for now and derive an effective single particle Hamiltonian

\begin{equation}
    H = -\frac{\hbar^2}{2m_{oxy}}\nabla^2+V(\mathbf{r}),
    \label{eq:OHam}
\end{equation}

where $m_{oxy}$ is the mass of an oxygen atom and $V(\mathbf{r})$ is the potential due to the surrounding (mostly amorphous) lattice.

The two most striking properties of a strongly coupled TLS, assuming the charge dipole framework is physically representational, are its ground to first excited state splitting $E_{01}$ and a strong electric dipole moment. Observed values of $E_{01}$  differ between qubit architectures: approximately $1$--$10$ GHz for transmons~\cite{Koch2007}, $4$--$5$ GHz for flux qubits~\cite{Lupascu2009} (although recent designs can tune this gap down to the MHz range~\cite{Schwarz2013}) and nominally $7$--$8$ GHz for phase qubits~\cite{Cole2010}. Dipole moment strengths also vary, but are usually on the order of $1 \; e\rm{\AA}$~\cite{Cole2010,Shalibo2010}. Whilst many other properties have been measured, this work will focus on obtaining respectable values for $E_{01}$ and dipole strength, assuming our model defect is embedded inside a fictitious phase qubit. When a representative example is required in this work, $E_{01}=8$ GHz and a dipole strength of $1.2 \; e\rm{\AA}$ will be used, to compare directly with a TLS studied in Ref. \onlinecite{Cole2010}.

A potential which represents the junction, requires a number of capabilities. It needs to describe interactions between atomic species (in this case, Al--Al, O--O and Al--O interactions), as well as many body interactions to be accurate as possible (which will be required to investigate quasi-degenerate states). As a complete description of many body interactions does not currently exist; potentials of this type tend to be empirically fitted to experimental data in order to obtain physico-chemical properties of a studied system as accurately as possible. The trade off here is that whilst any given potential may describe some properties of the system well because of certain fitting parameters, other properties may be well out of range as they were not included in the study that constructed the potential. Great care was taken to choose the best potential that accurately represented the junction, in this case the empirical Streitz-Mintmire potential~\cite{Streitz1994}, which describes a myriad of aluminium oxide properties over quite a large range of temperatures and pressures with high accuracy when compared against similar potentials. It was also chosen over simpler fixed-charge models~\cite{Catlow1982,Dienes1975} due to the complex geometry of the Josephson junction. A variable charge potential such as Streitz-Mintmire can capture the variable oxygen states when present in a predominantly metallic environment through the minimisation of the electrostatic term in the potential (Eq.\ \ref{eq:sm}). This capability is particularly important here, as our junction has two metal-oxide interfaces, and our TLS defects may reside close to these boundaries.

The Streitz-Mintmire potential is given by
\begin{equation}
\label{eq:sm}
V(\mathbf{r}) = E_{EAM}+\sum_{i}^{N}q_i\chi_i + \frac{1}{2}\sum_{i,j}^{N}q_{i}q_{j}V_{ij},
\end{equation}

where $q_{i,j}$ is the atomic charge, $\chi_i = \chi_i^0 + \sum_{j}Z_{j}(\left[j|f_i\right]-\left[f_i|f_j\right])$ and $V_{ij} = J_{i}^{0}\delta_{ij}+\left[f_i|f_j\right]$. $J_i^0$ is an empirical parameter known as ``atomic hardness'' or a self-Coulomb repulsion\cite{Rappe1991}, $\delta_{ij} = 1$ when $i=j$ and $\delta_{ij} = 0$ when $i\neq j$, and all summation is calculated for $N$ atoms of the target system. The square bracket notation represents Coulomb interaction integrals between valence charge densities and/or effective core charge densities and take the form \cite{Zhou2004}:

\begin{align}
\left[a|f_b\right] &= \int \frac{f_b(r_b,q_b)}{r_{av}}\,\mathrm{d}V_b\\
\left[f_a|f_b\right] &= \iint \frac{f_a(r_a,q_a)f_b(r_b,q_b)}{r_{vv}}\,\mathrm{d}V_a \mathrm{d}V_b
\end{align}

with $a=i,j; b=i,j; a\neq b$ in Eq.\ (\ref{eq:sm}). $\mathrm{d}V_{a,b}$ are integrating volume units. $r_{av}$ is therefore the center distance between atom $a$ and $\mathrm{d}V_b$, and $r_{vv}$ is the center distance between $\mathrm{d}V_a$ and $\mathrm{d}V_b$.

The first term in Eq.\ \ref{eq:sm}, $E_{EAM}$, does not depend on the partial charges $q_i$ and therefore describes a charge-neutral system, represented here with a quantum mechanical based empirical embedded atom model (EAM) for the Al-Al and Al-O interactions

\begin{equation}
E_{EAM} = \sum_{i}^{N}F_{i}\left[\rho_i\right]+\sum_{i<j}^{N}\phi_{ij}(r_{ij}),
\end{equation}

with $F_{i}\left[\rho_i\right]$ as the energy required to embed atom $i$ in a local electron density $\rho_i$, and $\phi_{ij}(r_{ij})$ describing the residual pair-pair interactions by way of Buckingham and Rydberg potentials
\begin{align}
\label{eq:smpair}
\phi_{ij}(r_{ij}) &= A\exp\left(-\frac{r_{ij}}{\rho}\right) \nonumber \\
&-B\left[1+C\left(\frac{r_{ij}}{r_0}-1\right)\right] \nonumber \\
&\times\exp\left[-C\left(\frac{r_{ij}}{r_0}-1\right)\right],
\end{align}
where $r_{ij}$ is the interatomic (Euclidean) distance between atoms $i$ and $j$, all other constants are listed in Table \ref{tab:smconsts}. Further formalisms and parameters can be found in Refs.\ \onlinecite{Streitz1994,Zhou2004}, implementation is also discussed in Ref.\ \onlinecite{Gale2003}.

\begin{table}[thb]
\caption{\label{tab:smconsts} Empirical constants for the Streitz-Mintmire pair potentials (Eq \ref{eq:smpair})~\cite{Streitz1994,Gale2003}.}
\begin{ruledtabular}
\begin{tabular}{ c|rrrrr }
Pair &
\multicolumn{1}{c}{A} &
\multicolumn{1}{c}{$\rho$} &
\multicolumn{1}{c}{B} &
\multicolumn{1}{c}{C} &
\multicolumn{1}{c}{$r_0$}  \\
\hline
Al-Al & 4.474755 & 0.991317 & 0.159472 & 5.949144 & 3.365875 \\
Al-O & 62.933909 & 0.443658 & 0.094594 & 9.985407 & 2.358570 \\
O-O & 3322.784218 & 0.291065 & 1.865072 & 16.822405 & 2.005092 \\
\end{tabular}
\end{ruledtabular}
\end{table}

Using this potential, a single body time-independent Hamiltonian is constructed using a $7$-point central difference method
%\begin{widetext}
%\begin{equation}
\begin{multline}
f^{\prime\prime}(x_0)=\\
\frac{2f_{-3}-27f_{-2}+270f_{-1}-490f_{0}+270f_{1}-27f_{2}+2f_{3}}{180h^{2}}\\
+\mathcal{O}(h^{6})
\end{multline}
%\end{equation}
%\end{widetext}
where $f_k=f\left(x_0+kh\right)$. The step size $h=0.1\;\rm{\AA}$ was found to be optimal for the grid range relevant to our purposes\cite{Mathews2004}. The resulting matrix is then diagonalized, obtaining eigenvectors and eigenenergies with precision better than $10$ kHz. In contrast, the energy scale for JJ defects observed in experiments is $\lesssim 10$ GHz~\cite{Neeley2008, Lupascu2009, Lisenfeld2010}.

The dipole element is computed using numerical integration of the ground- and first-excited states ($\psi_0$, $\psi_1$), where
\begin{equation}
    \wp_x = \iint \psi_0^*(x,y) x \psi_1(x,y) \,\mathrm{d}x\mathrm{d}y
    \label{eq:dipole}
\end{equation}
is an example of the dipole in the $x$ direction.

Relative differences in energy levels (i.e.\ energy splittings) are an important measure of the model. We therefore define a convention where $E_{ij} = E_j-E_i$, such that the ground state ($E_0$) to first excited state ($E_1$) energy splitting is defined as $E_{01}$. For comparison with experimental results, energy is expressed in frequency units throughout this paper.

\section{TLS Defects as Perturbed Bond Angles in a Lattice}\label{sec:bonds}

Consider the two simplest cases in Figure \ref{fig:defects} -- defect type A; where the aluminium--oxygen bond distance is shortened, forcing the oxygen to occupy two off axis positions, and defect type B; where the opposite occurs: aluminium--oxygen bond distance is lengthened, allowing two preferred oxygen positions on axis.

These two defect types can be modelled by solving our system hamiltonian (Eq.\ \ref{eq:OHam}) for three atoms: an oxygen with two aluminium atoms at a lattice coordinate apart, displaced toward and away from the oxygen. For example, corundum: the low pressure and temperature phase of aluminium oxide (also known as $\rm{\alpha}$-$\rm{Al_2O_3}$) has an Al--O bond distance of $\sim\!1.85\;\rm{\AA}$. If we define the oxygen position to be at an origin, the aluminium atoms can be considered as pairs ($x = -X, \: +X$ = \{$\pm 1.85 \; \rm{\AA}$\}) lying on a cardinal axes; which are identified throughout this paper as $\abs{X}$. Displacing $\abs{X}$ equidistantly from this origin (i.e.\ moving away from optimal crystalline configuration) will yield either an A or B type defect, depending on the direction of displacement.

%moved to fix float position
\begin{figure}[t]
  \centering
  \includegraphics[width=\columnwidth]{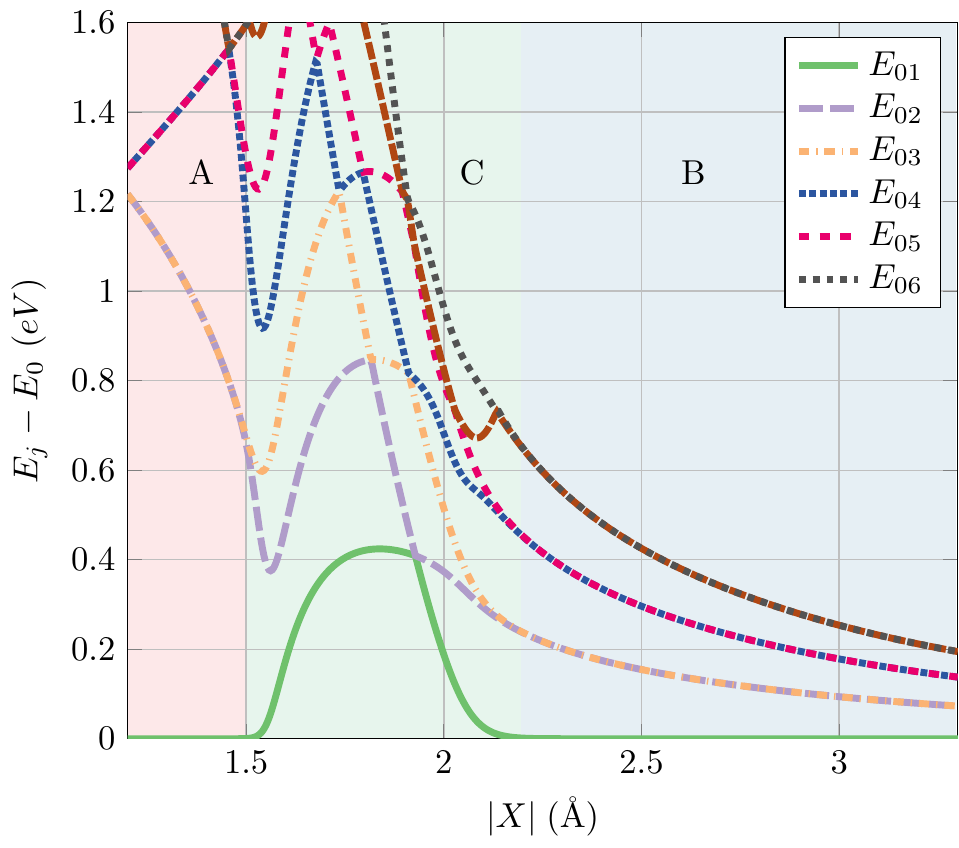}\\
  \caption{\label{fig:spectrum}(Color online) Eigenspectrum of a three atom system Al--O--Al, over a varying distance separation. Each excited state has been normalised with the ground state, which clearly shows two regions with a degeneracy at the lowest level. Section A is indicative of the A type defect (Figure \ref{fig:Atype}), section B of the B type (Figure \ref{fig:Btype}). An (an)harmonic crossover point is also extant, labelled as section C, which is approximately centered about the optimal Al--O bond distance of corundum ($1.85 \; \rm{\AA}$).}
\end{figure}

%moved to fix float position
\begin{figure}[b!]
\centering
\includegraphics[width=\columnwidth]{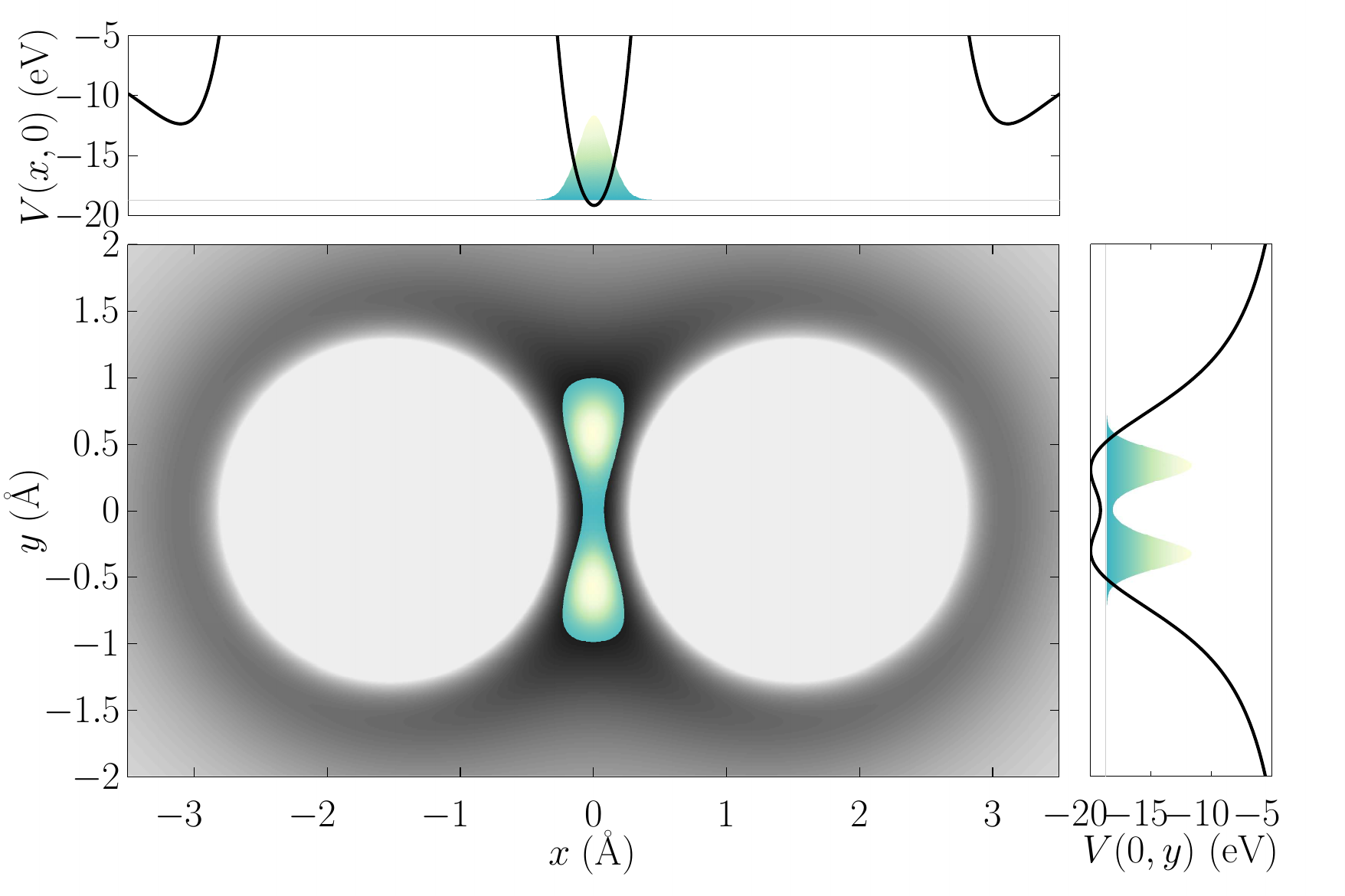}\\
\caption{\label{fig:Atype}(Color online) A top down view of the potential seen by an oxygen atom with two aluminium atoms closer than an appropriate lattice distance (truncated at $0$ eV, gray). The ground state eigenmode of the oxygen can be seen in the center (green/yellow). This configuration has a separation distance of $\abs{X} = 1.5 \; \mathrm{\AA}$ and is representative of an A type defect (see Figure \ref{fig:defects}). 1D potential values and projected wavefunctions, which are anchored to the ground state energy (thin gray line) are plotted in the outsets to better indicate the depth of the potential well.}
\end{figure}

An eigenspectrum of the six lowest energy levels of this system over a continuum of values in $\abs{X}$  is depicted in Figure \ref{fig:spectrum}. Each energy is measured relative to the ground state, which shows two particular regions where $E_{01}$ (green, solid) is quasi-degenerate (labelled sections A and B: both associated with the respective defect type). There exists a third (an)harmonic region (section C), which reaches a harmonic state at a separation distance of $\abs{X} \sim 1.85\;\rm{\AA}$: the optimal corundum Al--O bond distance. At this distance the spatial harmonic approximation holds and the oxygen can be considered to be localised.

%Eigenspct position

%moved to fix floats
\begin{figure}[b]
\centering
\includegraphics[width=\columnwidth]{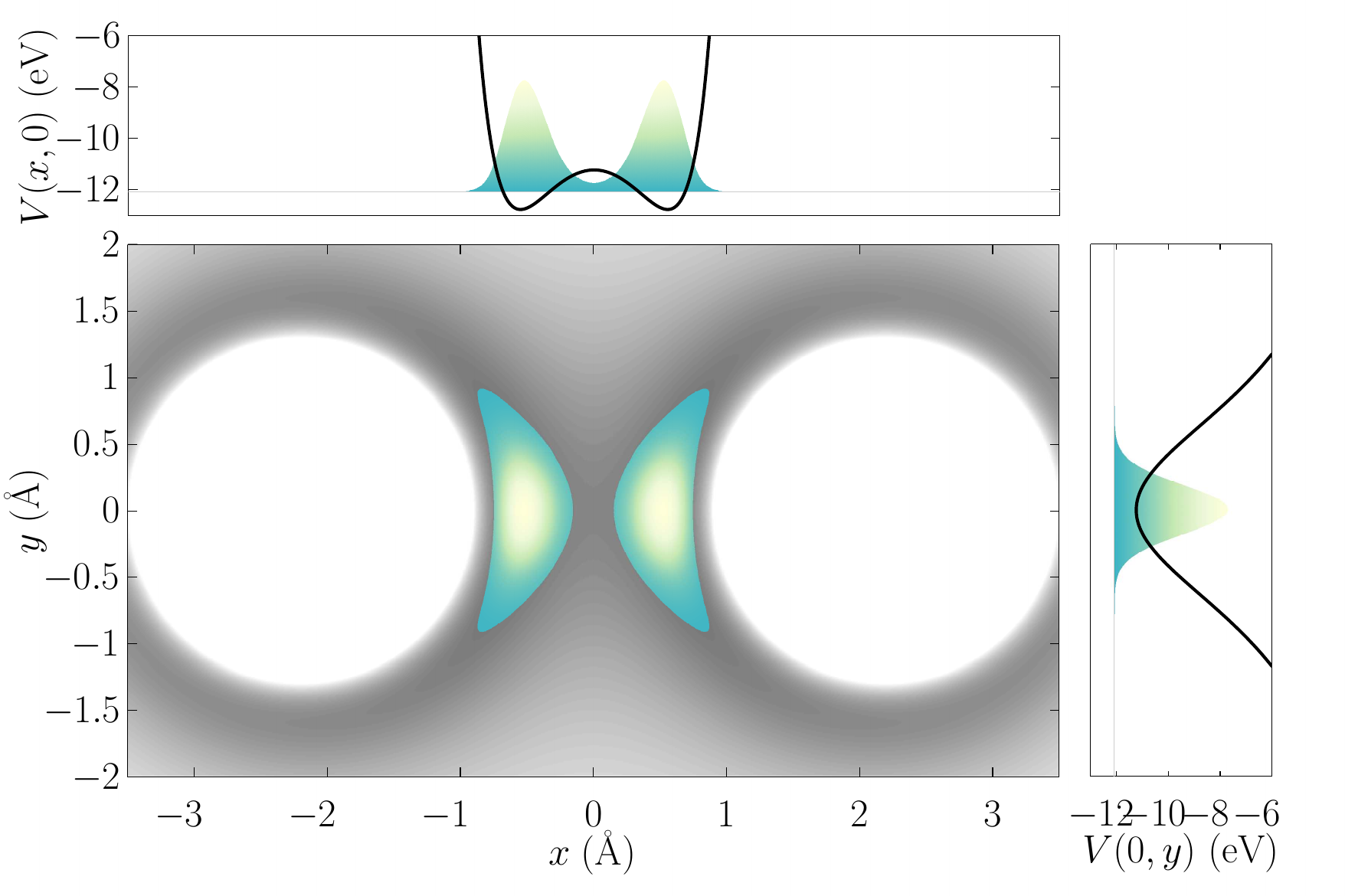}\\
\caption{\label{fig:Btype}(Color online) A top down view of the potential seen by an oxygen atom with two aluminium atoms further apart than an appropriate lattice distance (truncated at $0$ eV, gray). The ground state eigenmode of the oxygen can be seen in the center (green/yellow). This configuration has a separation distance of $\abs{X} = 2.2 \; \mathrm{\AA}$ and is representative of a B type defect (see Figure \ref{fig:defects}).  1D potential values and projected wavefunctions, which are anchored to the ground state energy (thin gray line) are plotted in the outsets to better indicate the depth of the potential well.}
\end{figure}

To investigate the potential landscape and the resultant oxygen wavefunction in sections A and B of Figure \ref{fig:spectrum}, we choose two separation distances: $\abs{X} = 1.5 \; \mathrm{\AA}$ (Figure \ref{fig:Atype}), which lies in the A type defect section, and $\abs{X} = 2.2 \; \mathrm{\AA}$ (Figure \ref{fig:Btype}), which exists in the B type defect section. Although the Al--O--Al chain is arranged in a line, we consider delocalisation of the oxygen in two dimensions so that both defect types can be identified on a continuum separation in $\abs{X}$ (as Figure \ref{fig:spectrum} depicts) rather than using two separate coordinate systems.

Both figures show a top down view of the potential exerted on the oxygen by the two aluminium atoms (gray), whose positions lie at the centre of the circles. We truncate potential energy values above $0$ eV for clarity, where energy is measured relative to the zero point set by the Streitz-Mintmire potential's electronegativity correction. The ground state wavefunction (green/yellow) indicates the spatial probability of the oxygen atom in these configurations. It can be clearly seen in Figure \ref{fig:Btype}, where the pair is far apart, two local minima exist in the form of rings around the base of each aluminium position: due to the fact that aluminium oxide is only partially covalent and mostly ionic.

Equivalences between the A and B type defects illustrated in Figure \ref{fig:defects} are clearly visible: Figure \ref{fig:Atype} showing a shortened bond length, causing an oxygen dipole perpendicular to the bond axis, and Figure \ref{fig:Btype} depicting a lengthened bond and an oxygen dipole parallel to the bond axis.

%position of Atype and BType

The functional form of the 1D potential (outset axes of Figures \ref{fig:Atype} and \ref{fig:Btype}, black; either $V(x,0)$ or $V(0,y)$) in the direction of the defect is a double well. Also depicted in these outsets are projected wavefunctions, which have been scaled for visual purposes, but anchored at the ground state energy (thin gray line). This representation is equivalent to the standard two level physics depiction of the TLS, and as such, potential offsets and tunneling matrix elements of our model may be estimated in this limit.

\section{TLS Defect Confined in Two Dimensions}\label{sec:2d}

The ideal case discussed in Section \ref{sec:bonds} ignores many real world complications, in particular any potential constraints from nearest neighbour atoms that undoubtedly surround the TLS. In an attempt to add complexity to the model gradually, we start with two additional aluminium atoms on the same plane as Figures \ref{fig:Atype} and \ref{fig:Btype}, confining the defect in the $\abs{Y}$ direction (i.e.\ $y = -Y, \: +Y$).

\subsection{Classifying Eigenspectrum Dynamics}

As the values of $\abs{X}$ or $\abs{Y}$ are altered, a complex interplay between the excited states of the model can result. Simple two-level degeneracy and harmonic states are no longer the only possibilities. To interpret what is occurring in a certain domain, we define a metric using the ground and four lowest excited state energies
\begin{equation}
\xi=\frac{E_{1}-E_{0}}{E_{2}-E_{0}}+\frac{E_{3}-E_{0}}{E_{4}-E_{0}}.
\end{equation}
This metric ranges from $0$ to $2$ and can give a qualitative understanding of the eigenspectrum of the defect.

\begin{figure}[htp]
\centering
\includegraphics[width=\columnwidth]{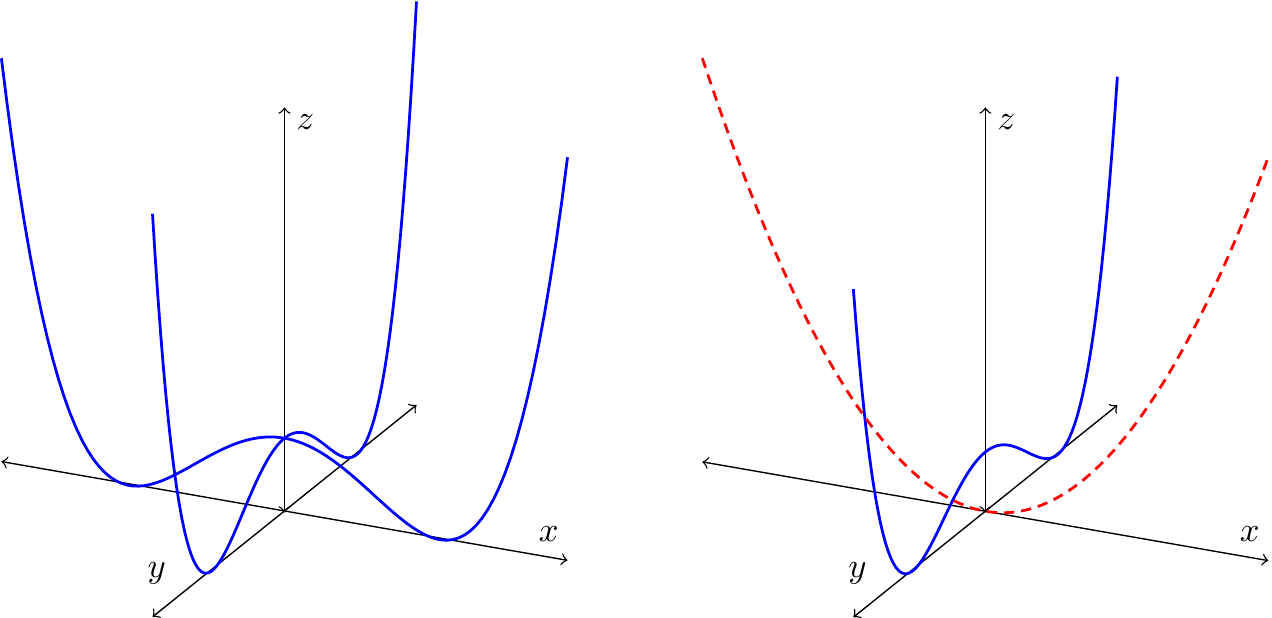}\\
\caption{\label{fig:mexhatproj}(Color online) 1D double wells (blue, solid) and harmonic wells (red, dashed) can be used to represent simple projections of a 2D potential onto the $x$ and $y$ axes. Left: two projected double wells is an example of a tetra-well. Right: a combination of one double well and a harmonic well reflects the hemi-tetra- case.}
\end{figure}

To begin we plot a phase space diagram akin to those introduced in Ref.\ \onlinecite{DuBois2013}, where $\xi$ is plotted as a function of the distance to the confining aluminium atoms ($\abs{X},\abs{Y}$). Each phase diagram is split into at least four domains, where the properties of these domains can be explained through the interplay of potential configuration and dipole alignment (discussed in Section \ref{sec:dipole}). Focusing for now on the influence of potential shape, the 2D potential can be approximated as two 1D potentials: one projected in the $x$ direction and the other along $y$. There are two relevant configurations, a set of two double wells (tetra-well) or a set of a double and harmonic well (hemi-tetra-well); which are both illustrated in Figure \ref{fig:mexhatproj}. It is clear from the outset potential projections of Figures \ref{fig:Atype} and \ref{fig:Btype} that both A and B type defects reside in hemi-tetra-wells.

\begin{figure}[bhp]
  \centering
  \includegraphics[width=0.9\columnwidth]{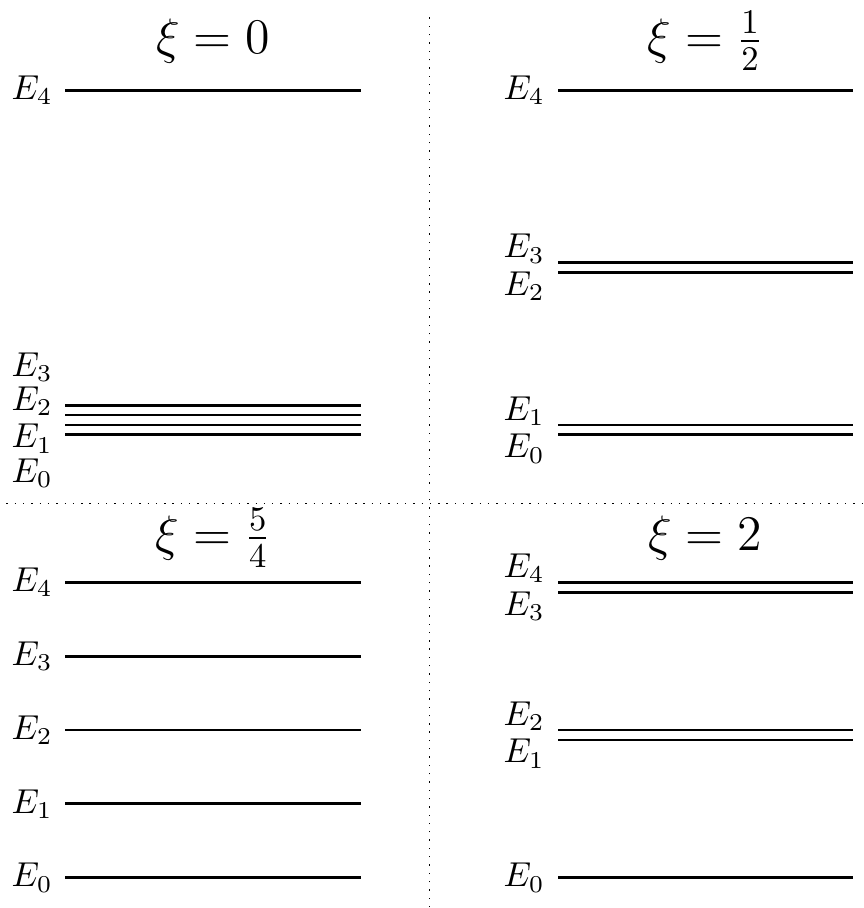}\\
  \caption{\label{fig:ximetric}Energy level representation of the lowest five eigenenergies of a candidate defect and their associated $\xi$ value. $\xi=0$: the tetra-well domain, $\xi=1/2$: the hemi-tetra-well domain, $\xi=5/4$: the harmonic region and $\xi=2$: the unique ground state region. The bifurcation/transition region, $\xi=1$ is not clearly defined thus an example is not shown here.}
\end{figure}

The $\xi$ metric is capable of identifying the tetra- ($\xi=0$) and hemi-tetra- ($\xi=1/2$) domains, as well as harmonic ($\xi=5/4$) and unique ground state (rotationally symmetric, Mexican hat-like) ($\xi=2$) regimes and finally the location of bifurcations or transitions ($\xi=1$). Figure \ref{fig:ximetric} shows the corresponding layouts of each interplay. It is worth noting that $\xi=3/2$ can also be considered harmonic for the lowest three levels.

\subsection{Visualising Phase Space and Identifying TLSs}\label{sec:phasespace}

\begin{figure}[t]
\includegraphics[width=\columnwidth]{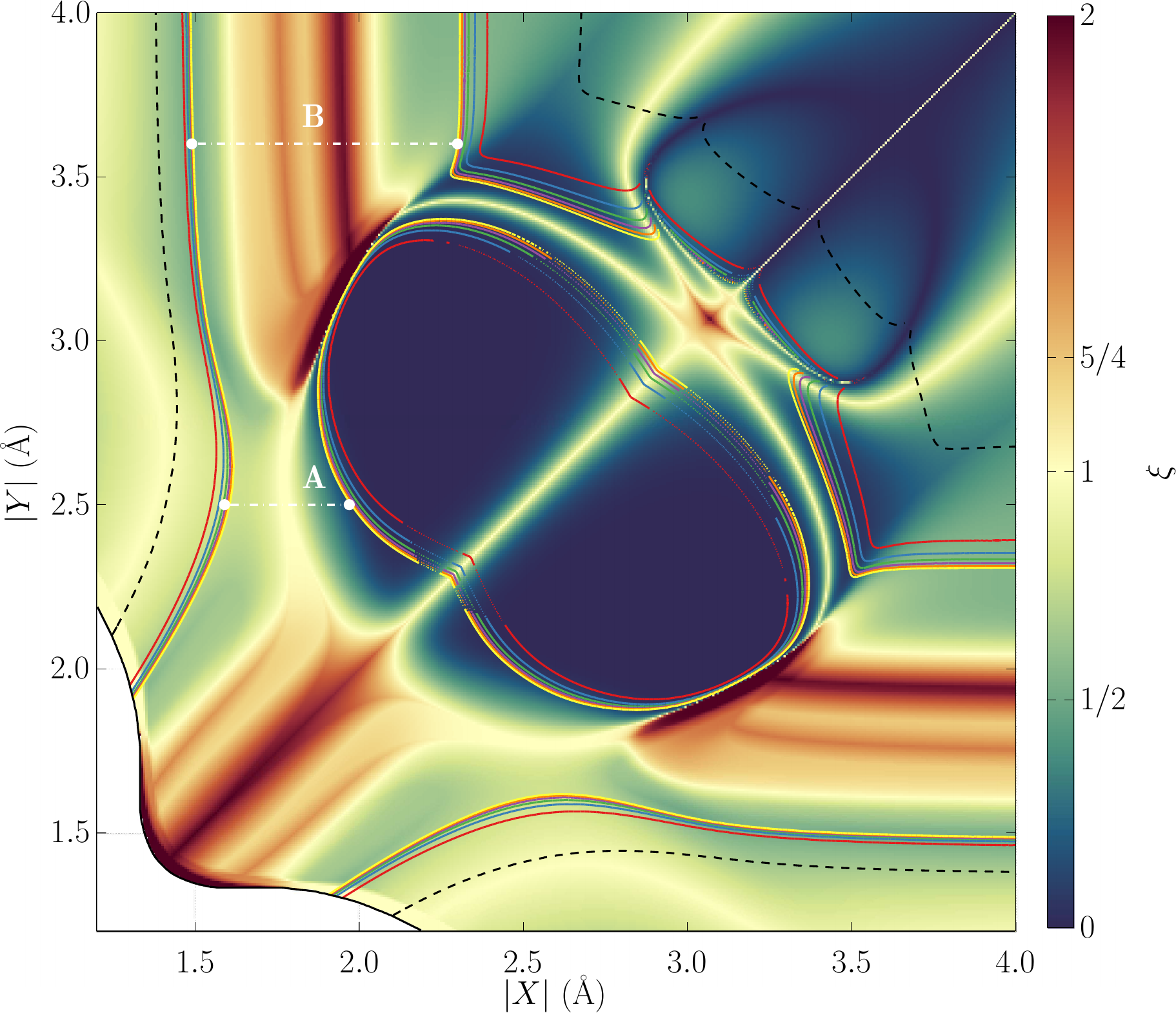}\\
\caption{\label{fig:xiunbound}(Color online) Map of the $\xi$ metric of the delocalized oxygen 2D model. The $\abs{X}$ and $\abs{Y}$ axes represent aluminium pair position separations. Black, dashed contour lines represent a minimum resolvable energy splitting of $10$ kHz.  The white (blank) section indicates where the aluminium atoms are so close, the oxygen confinement region no longer exists. Overlayed contour lines corresponding to $E_{01} = 0.5\!-\!10$ GHz (red to yellow) are comparable to existing experimental qubit results. Cases where quad-degeneracy exists are denoted as dotted rather than solid contours. Two traces (white, dash-dotted lines) are also depicted. The first, at $\abs{Y} = 2.5 \; \rm{\AA}$ (labelled A) is plotted in Figure \ref{fig:tetraspectrum}. In contrast, the trace at $\abs{Y} = 3.6 \; \rm{\AA}$ (labelled B) yields an equivalent eigenspectrum to the 1D case in Figure \ref{fig:spectrum}. }
\end{figure}

Using this $\xi$ metric and varying the values of both $\abs{X}$ and $\abs{Y}$ generates Figure \ref{fig:xiunbound}, a phase map of the interplay of low energy states of the oxygen atom confined in two dimensions by aluminium atoms. The $x$ and $y$ axes show the $\abs{X}$ and $\abs{Y}$ pair separation distances respectively over a range of $1.85\!-\!4 \: \mathrm{\AA}$ and the phase colour indicates which $\xi$ region a particular configuration exists in.

%position of xiunbound

Regions that are deemed to be unimportant when searching for TLS behaviour are those where $\xi \geq 1$, as these correspond to eigenspectra which do not possess a (quasi-)doubly degenerate ground state.

As discussed in Section \ref{sec:bonds}, in the one dimensional confinement case, both A and B type defects exist in a hemi-tetra-well ($\xi = 1/2$). This is also the case for the two dimensional landscape -- contour lines corresponding to $E_{01} = \{0.5, 2, 4, 6, 8, 10\}$ GHz (red--yellow) on Figure \ref{fig:xiunbound} show the configuration properties that result in TLS $E_{01}$ splittings in various qubit architectures described in Section \ref{sec:methods}. For example: anywhere an orange, $8$ GHz line exists on this phase map, the $\abs{X}, \abs{Y}$ coordinates of the model generate a cluster configuration that yields the same $E_{01}$ observed in phase qubit experiments~\cite{Cole2010}.

Many of these lines lie completely within a $\xi = 1/2$ (green) region. Consider the contour set on the left of Figure \ref{fig:xiunbound} where $\abs{X} \simeq 1.5 \; \rm{\AA}$: Section \ref{sec:bonds} states this distance is indicative of an A type defect. The addition of the $\abs{Y}$ aluminium pair does little to perturb the potential at larger distances ($\abs{Y} \gtrsim 3 \; \rm{\AA}$) and causes only slight deformations at smaller distances ($2 \lesssim \abs{Y} \lesssim 3 \; \rm{\AA}$). As the phase space is symmetric about $\abs{X} = \abs{Y}$, A type defects exist at the bottom of the plot where $\abs{Y} \simeq 1.5 \; \rm{\AA}$ as well.

B type defects also exist in a $\xi = 1/2$ region, when $\abs{X}$ or $\abs{Y} \simeq 2.4 \; \rm{\AA}$. The orthogonal pair separation distance is at least $1\; \rm{\AA}$ larger than the defect pair in these configurations and therefore have no bearing on the potential seen by the oxygen. Complications arise when the orthogonal pair is closer and begins to confine the defect. To explain this response, a better understanding of the domains of the map is required.

%moved to fix flats

\subsection{Analysis of Phase Space Domains}\label{sec:phaseanalysis}

The case of $\xi = 0$ is particularly interesting in this scenario: a tetra-well region, causing a quad degeneracy in the ground state. In this region, the characteristics of a B type defect are strongly modified. To understand why, we first explain the large rounded $\xi = 0$ domains in the centre of Figure \ref{fig:xiunbound}'s phase space.

\begin{figure}[thp]
  \centering
  \includegraphics[width=\columnwidth]{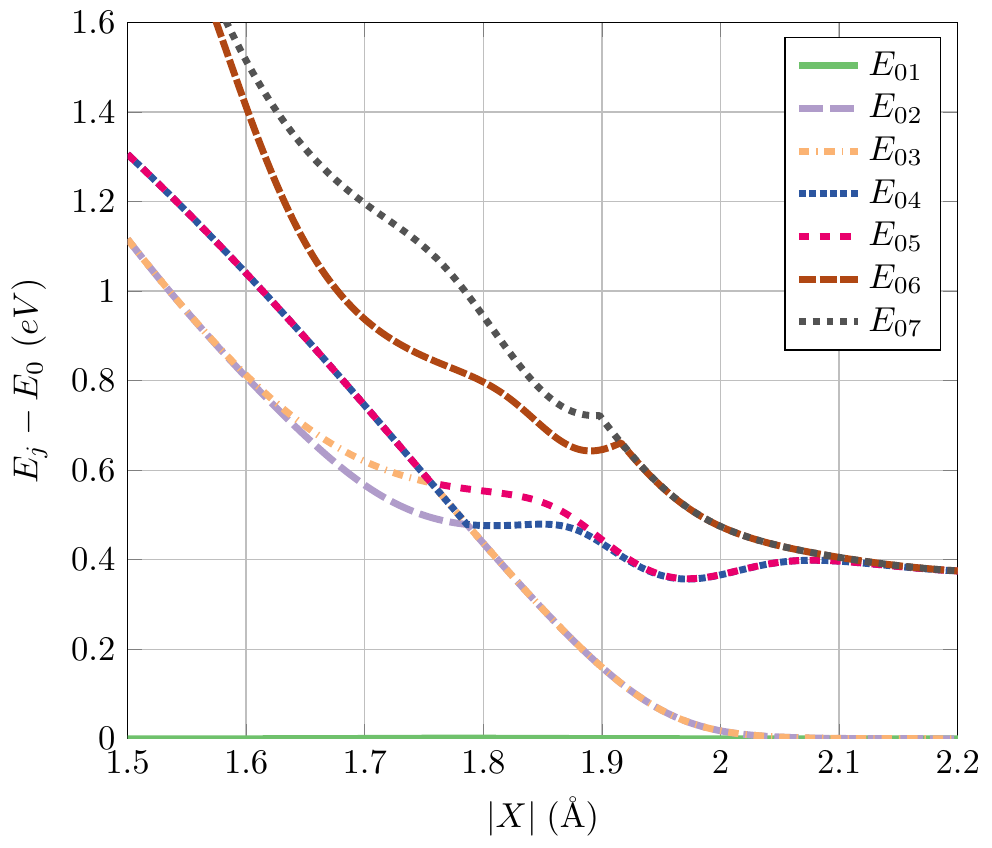}\\
  \caption{\label{fig:tetraspectrum}(Color online) Eigenspectrum of a 2D TLS: an oxygen atom caged with four aluminum atoms. Each excited state has been normalised with the ground state. The confinment pair $\abs{Y}$ is held at a constant distance separation of $2.5 \; \rm{\AA}$ and the $\abs{X}$ range shows how higher eigenvalues behave as the confinement atoms force the defect into a tetra-well regime. Figure \ref{fig:xiunbound} depicts this data in terms of $\xi$ (white dash-dotted trace labelled A). }
\end{figure}

Tetra-well domains exist when the confinement potential on the oxygen consists of two double wells (see left plot in Figure \ref{fig:mexhatproj}. This phenomena emerges when both the defect pair and the confining pair of atoms are close enough to interact. Consider an A type defect with a defect pair in $\abs{X}$ and a confining pair at a constant value of $\abs{Y} = 2.5 \; \rm{\AA}$. If $\abs{X}$ is varied from an initial value of $1.6 \; \rm{\AA}$ to $2 \; \rm{\AA}$, this extension spans phase space from one point where $E_{01} = 8$ GHz to another. This separation is traced on Figure \ref{fig:xiunbound} (white dash-dotted line labelled A), where the defect visibly moves from a hemi-tetra- regime ($\xi = 1/2$) and goes through a transition region before reaching the tetra-well regime ($\xi = 0$). Figure \ref{fig:tetraspectrum} depicts the eigenspectrum of this trace. This is in contrast to larger confining pair values such as $\abs{Y} = 3.6 \; \rm{\AA}$ (also traced on Figure \ref{fig:xiunbound}: white dash-dotted line labelled B) where the eiginspectrum is largely unchanged from the one dimensional case presented in Figure \ref{fig:spectrum}.

As the $\abs{X}$ separation distance is increased, the effect on $E_{01}$ is negligible on the scale of the higher energy levels (although differs greatly on the TLS splitting energy scale). However, the degenerate pair $E_{23}$ rapidly shifts from a level much higher than ground to quasi-degenerate at ground. Complete quad-degeneracy is not always extant in the tera-well regime, configurations of $\abs{X}$ and $\abs{Y}$ in these domains usually have two quasi-degenerate pairs which are still observable, akin to the hemi-tetra- domains, with the difference \textit{between} the pairs approaching the difference \textit{of} the pairs: $E_{01} = E_{23} \approx E_{12}$ ($1.95 \lesssim \abs{X} \leq 2 \; \rm{\AA}$ in Figure \ref{fig:tetraspectrum} for example).

When considering the influence of higher lying excited states, its important to keep the fundamental energy scales of the problem in mind. In typical qubit experiments, the superconducting properties of the device put a rigorous upper limit on TLS frequencies of interest. At frequencies greater than approximately $100$ GHz, there is enough energy to dissociate Cooper-pairs and therefore it can be viewed as an operational upper bound for Josephson junction devices (typical operating frequencies are however device specific, and are much lower in practice). For much of the tetra-well domain $E_{12} \geq 100$ GHz and consequently can effectively be ignored, the system can be considered as a two-level defect even in this quad degeneracy domain.

The B type defects' sudden behaviour change as $\abs{X}\rightarrow\abs{Y}$ is also caused by this response (see Figure \ref{fig:xiunbound}). Confinement pairs start interacting with the defect, causing $E_{23}$ to approach the value of $E_{01}$ (again, $1.95 \lesssim \abs{X} \leq 2 \; \rm{\AA}$ in Figure \ref{fig:tetraspectrum} depicts this phenomena). The $\xi$ metric does not clearly differentiate between two quasi-degenerate pairs that are marginally separated and two pairs that are actually degenerate.

If however, $E_{12} < 100$ GHz, higher lying energy states must still be considered and the model exhibits true quad degenerate behaviour. Regions in which this occurs are denoted in Figure \ref{fig:xiunbound} as dotted contours, which over the entirety of phase space are extremely rare -- which suggests a reason why quad-level systems are yet to be experimentally observed.

The final domain yet to be discussed on Figure \ref{fig:xiunbound} is the upper right hand corner where both $\abs{X}$ and $\abs{Y}$ are large. This region is tetra-well dominated but can be considered as a region where the TLS model breaks down. Each of the four potential minima exist localised about the four confining aluminium atoms and as such should not be considered as TLS candidates.

\section{TLS Defect Confined in Three Dimensions}\label{sec:tls}

\begin{figure}[bp]
  \centering
  \includegraphics[width=0.7\columnwidth]{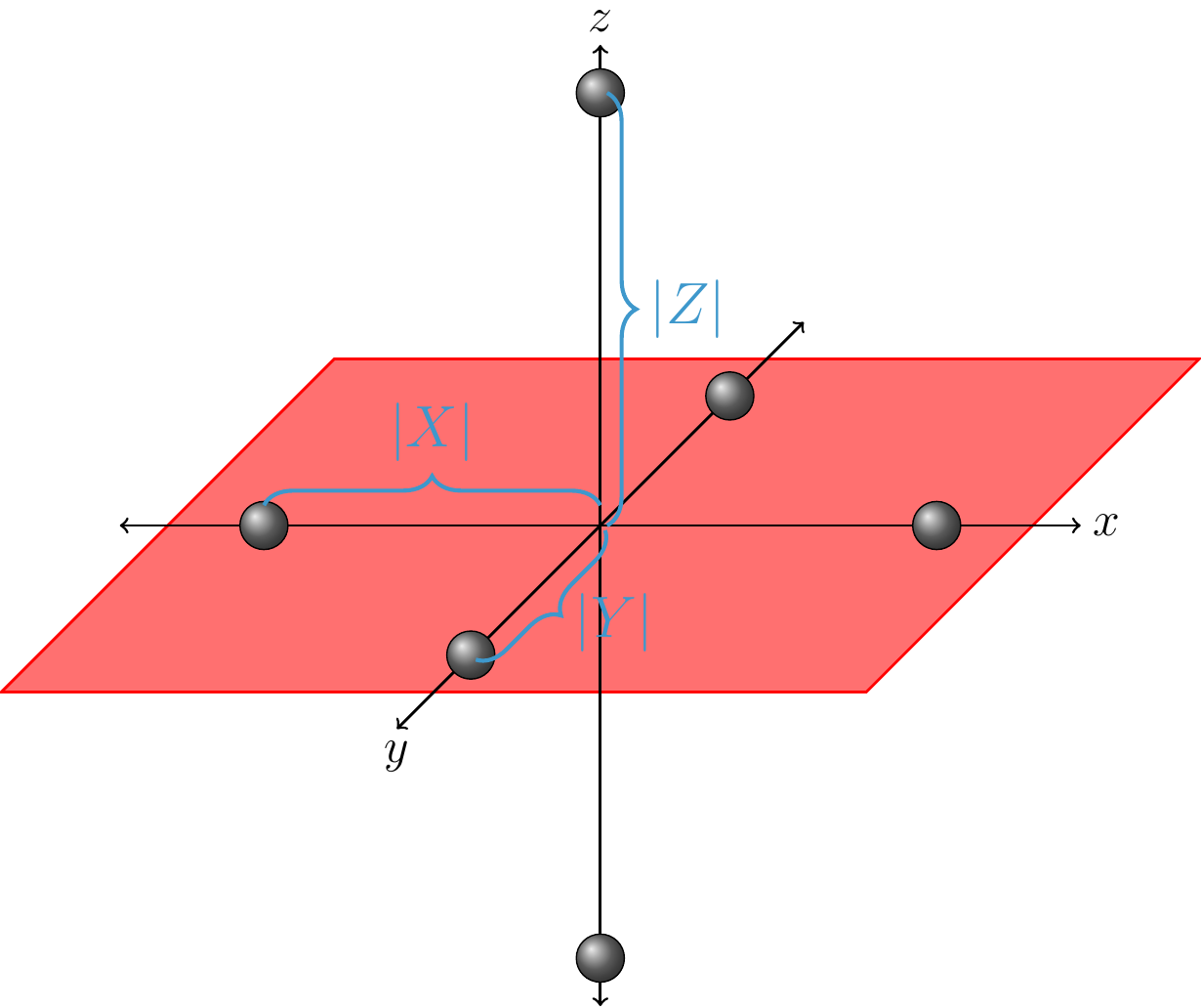}\\
  \caption{\label{fig:cage}(Color online) Illustration of an idealised cluster producing a void in 3D. Six cage aluminium atoms (gray) sit in pairs on each cardinal axis, equidistant from the origin. Separation distances $\abs{X}$ and $\abs{Y}$ are labeled for reference (see text). The plane (red) at $z=0$ is a representation of the 2D delocalisation of an oxygen atom.}
\end{figure}

\begin{figure}[t]
\centering
  \includegraphics[width=\columnwidth]{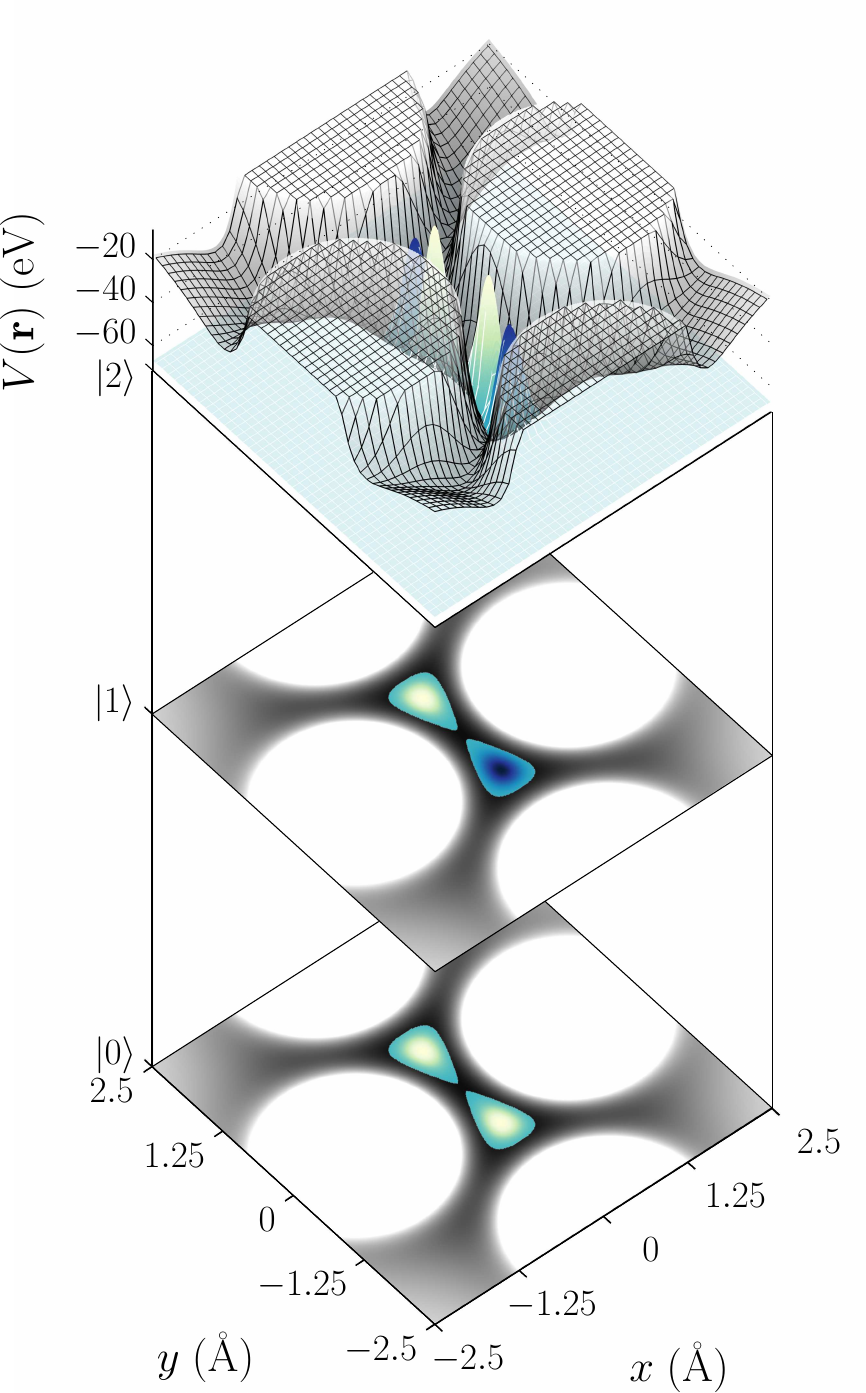}\\
\caption{\label{fig:wfstackA}(Color online) Cage potential and the lowest three eigenfunctions of a cluster with the values $\abs{X}=1.53, \: \abs{Y} = 2.52, \: \abs{Z} = 2.25 \; \mathrm{\AA}$. The top image in the stack is presented with the apparent `depth' of the potential well on the $z$-axis and the second excited state $\ket{2}$ scaled accordingly. Ground $\ket{0}$ and first excited $\ket{1}$ states are displayed in a projected representation underneath. This cluster configuration is representative of an A type defect (see Figure \ref{fig:Atype}); with a ground to first excited state splitting of $E_{01} = 8.1$ GHz and a ground to second excited state splitting of $E_{02} = 202.2$ THz.}
\end{figure}

\begin{figure}[t]
\centering
  \includegraphics[width=\columnwidth]{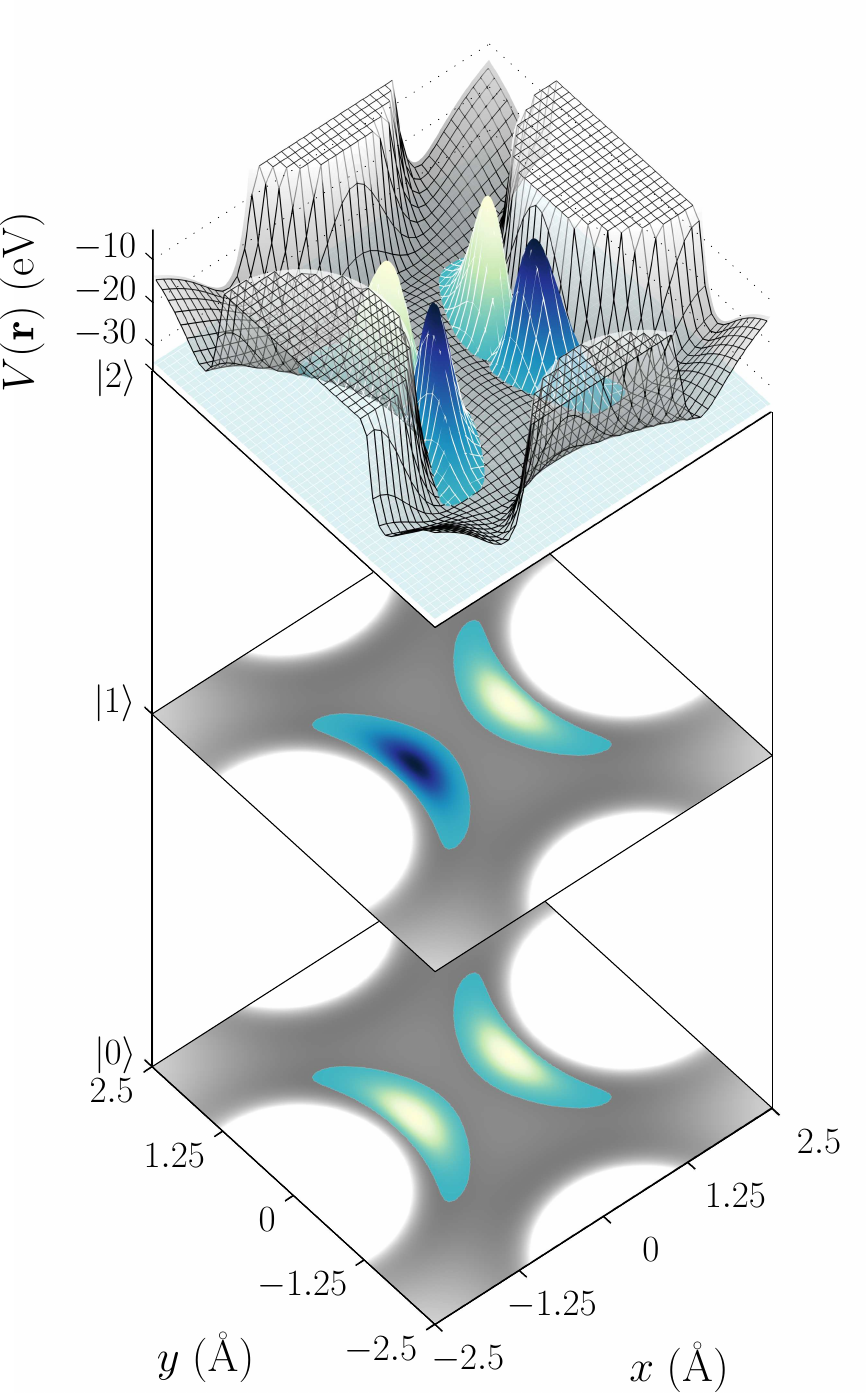}\\
\caption{\label{fig:wfstack}(Color online) Cage potential and the lowest three eigenfunctions of a cluster with the values $\abs{X}=2.38, \: \abs{Y} = 3.19, \: \abs{Z} = 2.75 \; \mathrm{\AA}$. The top image in the stack is presented with the apparent `depth' of the potential well on the $z$-axis and the second excited state $\ket{2}$ scaled accordingly. Ground $\ket{0}$ and first excited $\ket{1}$ states are displayed in a projected representation underneath. This cluster configuration is representative of a B type defect (see Figure \ref{fig:Btype}); with a ground to first excited state splitting of $E_{01} = 8.4$ GHz and a ground to second excited state splitting of $E_{02} = 36.3$ THz.}
\end{figure}

To understand the properties of a delocalised oxygen, we have considered confining aluminium atoms in both a line and in a plane. In reality, aluminium atoms will surround the oxygen in all three dimensions. Two more confining atoms are therefore added into the system in the $z$ direction labelled as $\abs{Z}$. Interactions with these atoms in the third dimension are now considered, although the model is still two dimensional (i.e.\ $2\!+\!1$D); thus oxygen continues to be confined to the $xy$ plane. An illustration of this cluster configuration is displayed in Figure \ref{fig:cage} and representations of the cage potential and oxygen wavefunctions are shown in Figures \ref{fig:wfstackA} and \ref{fig:wfstack} for A and B type defects respectively.

\begin{figure}[t]
  \centering
  \includegraphics[width=\columnwidth]{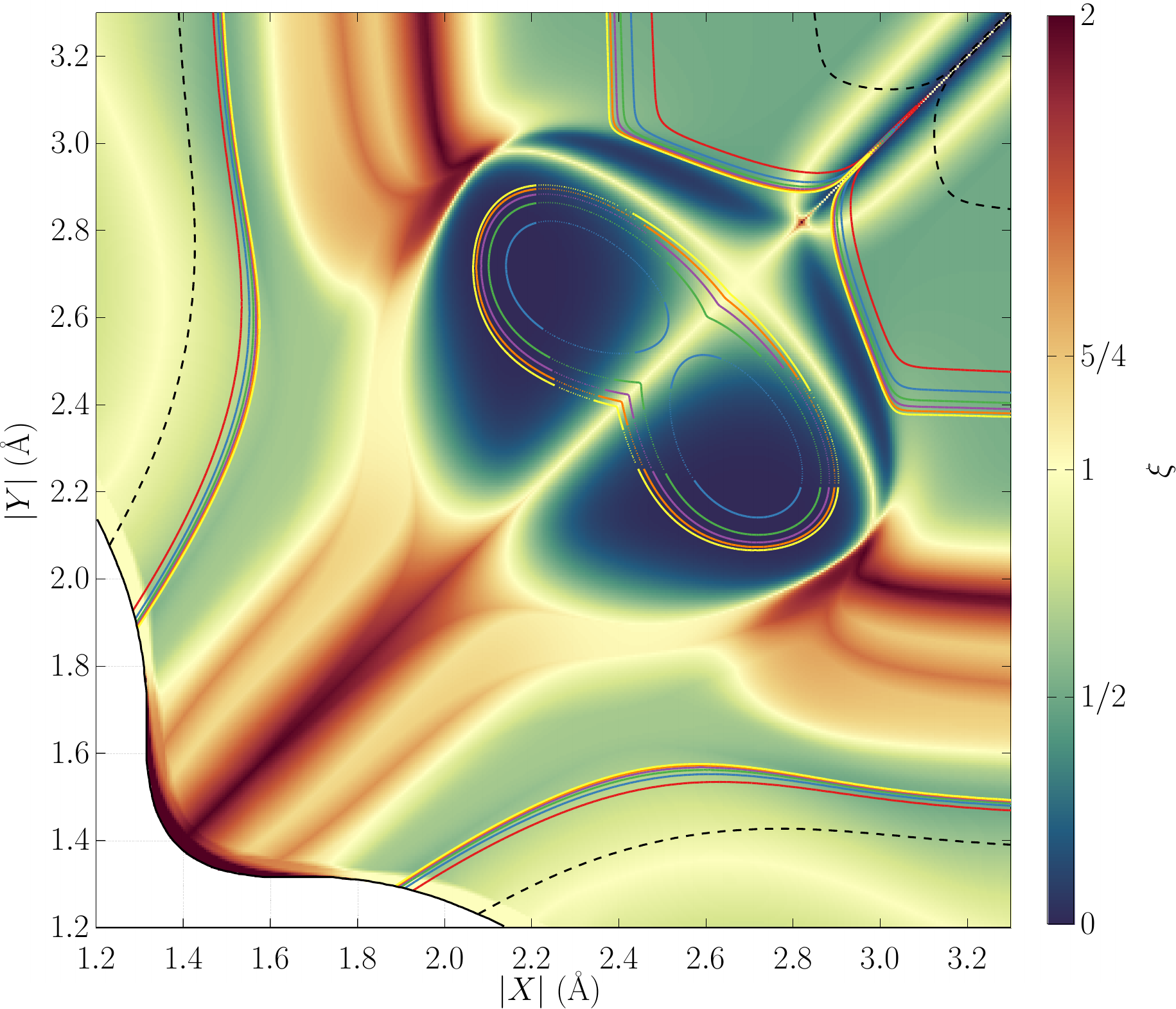}\\
  \caption{\label{fig:xi275}(Color online) Map of the $\xi$ metric of the delocalized oxygen model confined in two dimensions. The $\abs{X}$ and $\abs{Y}$ axes represent aluminium pair positions with $\abs{Z} = 2.75 \; \rm{\AA}$. Black, dashed contour lines represent a minimum resolvable energy splitting of $10$ kHz. The white (blank) section indicates where the aluminium atoms are so close, the oxygen confinement region no longer exists. Overlayed contour lines corresponding to $E_{01} = 0.5\!-\!10$ GHz (red to yellow) are comparable to existing experimental qubit results. Cases where quad-degeneracy exists are denoted as dotted rather than solid contours.}
\end{figure}

\begin{figure}[t]
  \centering
  \includegraphics[width=\columnwidth]{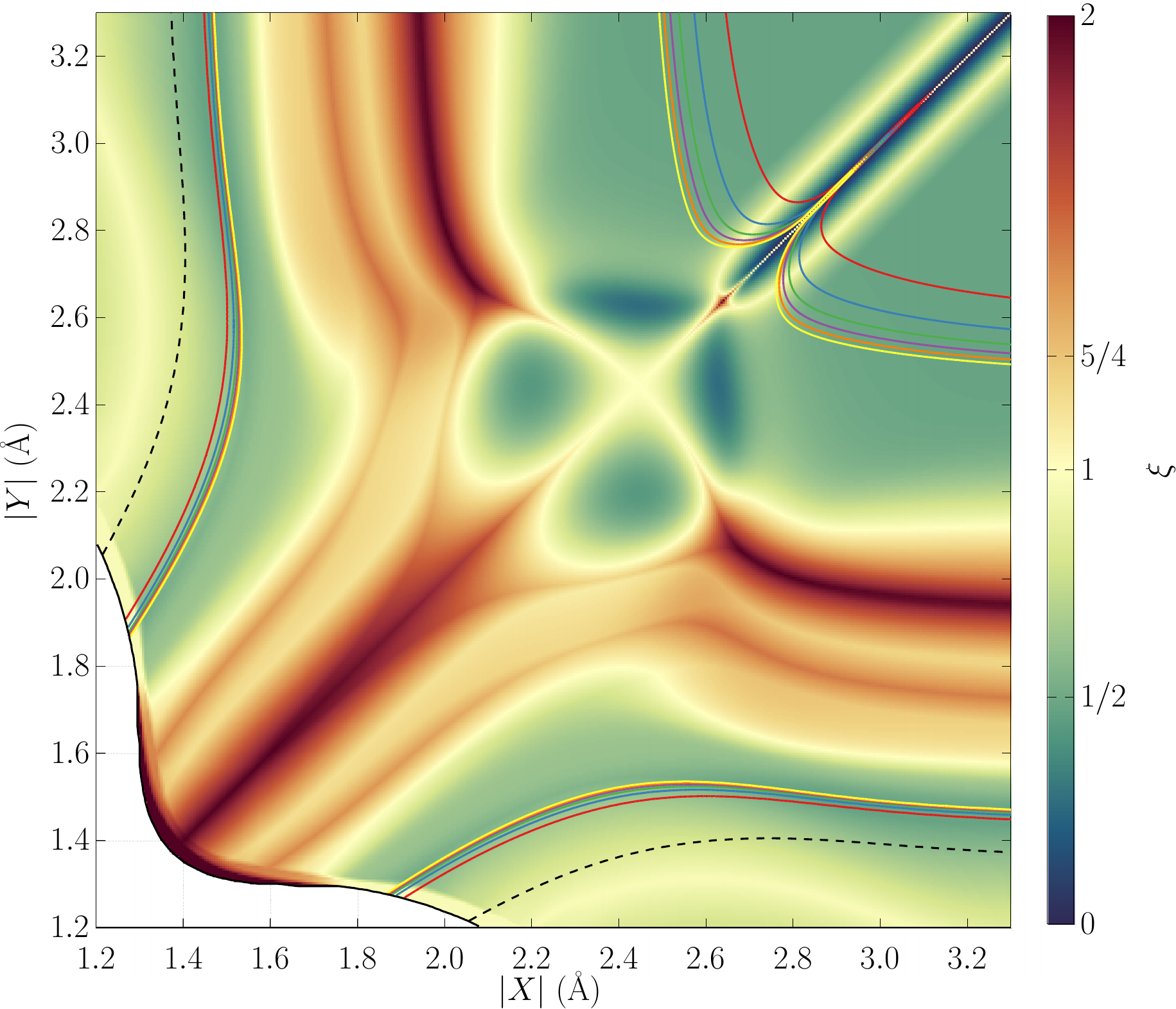}\\
  \caption{\label{fig:xi225}(Color online) Map of the $\xi$ metric of the delocalized oxygen model confined in two dimensions. The $\abs{X}$ and $\abs{Y}$ axes represent aluminium pair positions with $\abs{Z} = 2.25 \; \rm{\AA}$. Black, dashed contour lines represent a minimum resolvable energy splitting of $10$ kHz. The white (blank) section indicates where the aluminium atoms are so close, the oxygen confinement region no longer exists. Overlayed contour lines corresponding to $E_{01} = 0.5\!-\!10$ GHz (red to yellow) are comparable to existing experimental qubit results.}
\end{figure}

The selection of a fixed $\abs{Z}$ distance changes the phase landscape in a manner that can be qualitatively extrapolated between two arbitrary values even a few angstroms apart. Values of $\abs{Z} = 2.75 \; \rm{\AA}$ (Figure \ref{fig:xi275}) and $\abs{Z} = 2.25 \; \rm{\AA}$ (Figure \ref{fig:xi225}) have been chosen to analyse in detail.

TLS behaviour can be observed on both maps and each value of $\abs{Z}$ has been selected based on model parameters. Oxygen confinement occurs for $\abs{Z}$ values lower than $2.25 \; \mathrm{\AA}$ (i.e.\ $E_{01} \gg 100$ GHz). $\abs{Z}$ values larger than $2.75 \; \mathrm{\AA}$ show similar phase behaviour to that of Figure \ref{fig:xiunbound}, which in completely unbound in $z$. Large $\abs{Z}$ separation distances also decrease the validity of the $2\!+\!1$D model, in addition: the radial distribution analysis in Ref.\ \onlinecite{DuBois2013} suggests large separation distances for nearest neighbour atoms have a low probability of occurrence.

As the pair separation distance $\abs{Z}$ decreases, the tetra-well ($\xi=0$) regimes diminish in size and no longer exhibit TLS behaviour. This suggests that quad-degenerate defects, whilst quite rare in phase space already, are extremely rare in reality. For one to exist in a junction, the amorphous layer would have to be disordered in such a way that an oxygen atom's nearest neighbour atom pair exists at a distance greater that $\sim 3 \; \rm{\AA}$ along one orthogonal basis vector. %An investigation into the ramifications of stoichiometry and density of the amorphous layer will be presented in future work.

Whilst this configuration of six cage aluminium atoms on cardinal axes around a central oxygen is still an idealised system, Figure \ref{fig:xi225} is confined in all three spatial dimensions with pragmatic distances and is therefore considered to be the most `realistic' representation of the TLS phase space for this model published here. TLS candidates lie well within hemi-tetra- ($\xi=1/2$) domains, are not mired by higher energy level complexities (i.e.\ quad-quasi-degeneracies) and can be clearly identified as A type and B type defects separated by an (an)harmonic boundary.

Phase space is also dominated on this map with $\xi=5/4$ (harmonic) and $\xi=2$ (unique ground state) domains where the oxygen atom can be considered under the spatial harmonic approximation (i.e.\ not delocalised). This is significant, as TLS observations in experiments are not statistically dense. We expect few defects in the junction compared to the number of atoms extant.

\section{Charge Dipoles}\label{sec:dipole}

\begin{figure}[thp]
  \centering
  \includegraphics[width=\columnwidth]{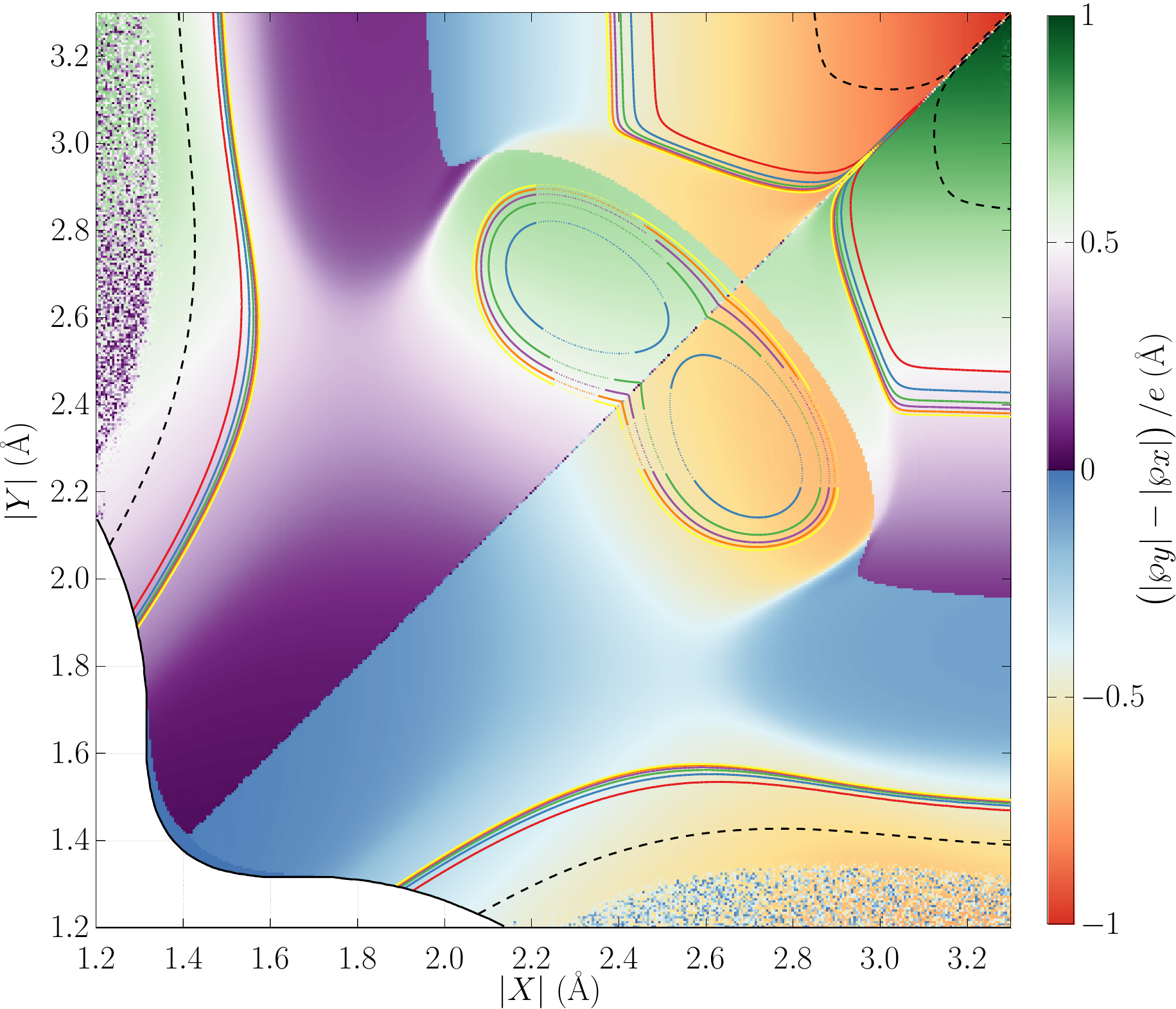}\\
  \caption{\label{fig:dipole275}(Color online) The difference between the absolute dipole moment (in $x$- and $y$-directions) over the same range for $\abs{Z} = 2.75 \; \rm{\AA}$. We see either $\abs{\wp_x}$ (red) or $\abs{\wp_y}$ (green) dominated behaviour in the tetra- and hemi-tetra- domains but none in other regions.}
\end{figure}

\begin{figure}[thp]
  \centering
  \includegraphics[width=\columnwidth]{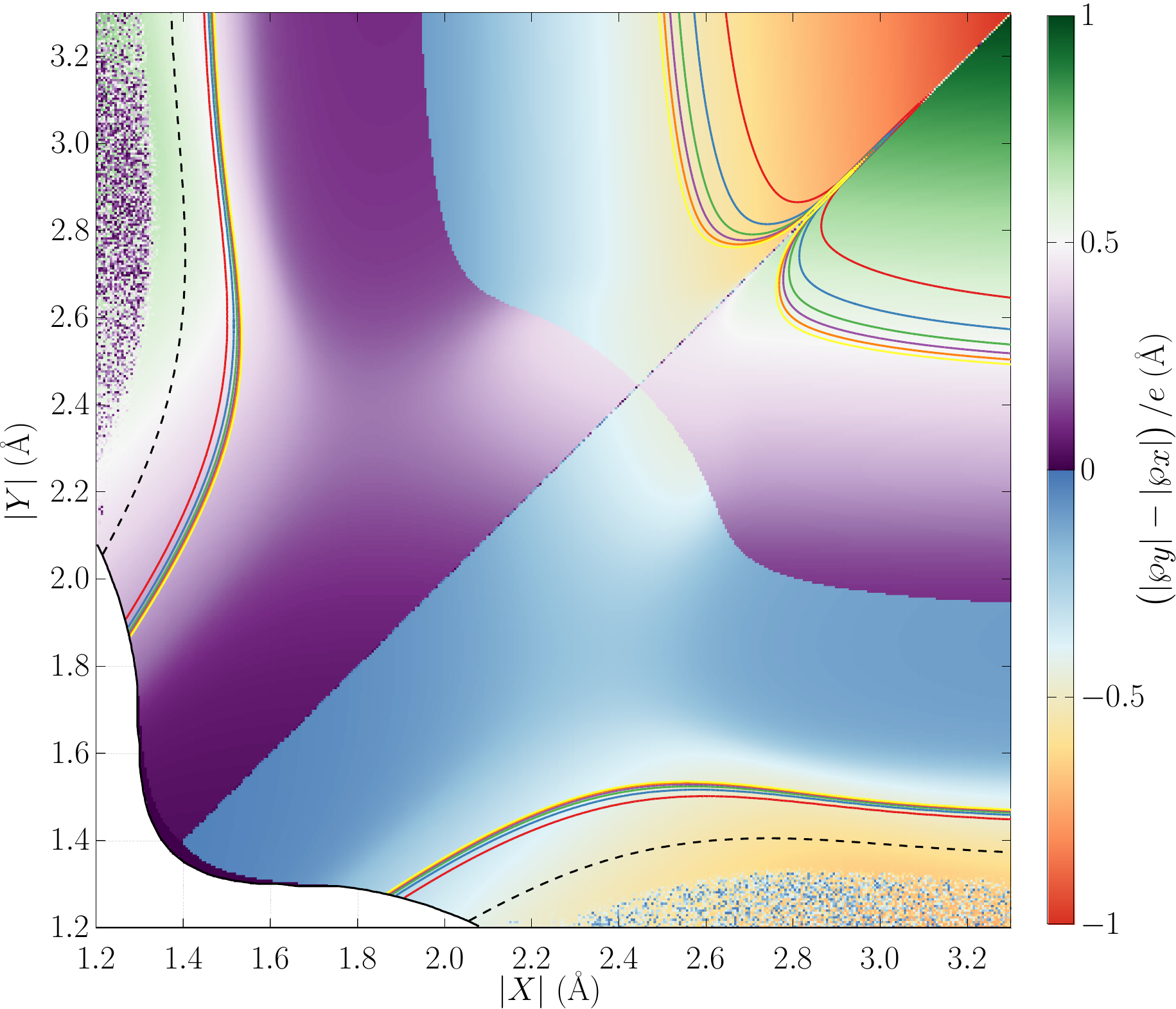}
  \caption{\label{fig:dipole225}(Color online) The difference between the absolute dipole moment (in $x$- and $y$-directions) over the same range for $\abs{Z} = 2.25 \; \rm{\AA}$. We see either $\abs{\wp_x}$ (red) or $\abs{\wp_y}$ (green) dominated behaviour in the tetra- and hemi-tetra- domains but none in other regions.}
\end{figure}

As stated in Section \ref{sec:methods}, another important experimentally measurable property of the TLS is its strong electric dipole moment. Using the same $\abs{X}, \: \abs{Y}$ and $\abs{Z}$ parameters from the phase maps in the previous section, the dipole elements in the $x$ direction $\wp_x$, and the $y$ direction $\wp_y$ can be calculated via Equation \ref{eq:dipole}. Figure \ref{fig:dipole275} shows the same phase space as Figure \ref{fig:xi275}, where $\abs{Z} = 2.75 \; \rm{\AA}$, and Figure \ref{fig:dipole225} matches Figure \ref{fig:xi225}, where $\abs{Z} = 2.25 \; \rm{\AA}$. The colourmap for both figures now represent dipole strengths of each cluster configuration rather than energy level splittings (represented through the $\xi$ metric). These computed dipole moments correspond well to observed values, assuming $\mathcal{O}\left(\rm{nm}\right)$ junction widths~\cite{Martinis2005, Cole2010}.

% Position of dipole225

The dipole moments are presented as $\left(\abs{\wp_y}-\abs{\wp_x}\right)/e$ rather than separate plots because the dipole elements are discontinuous at the bifurcation points (i.e.\ when $\abs{\wp_x}>0, \abs{\wp_y}=0$ and \textit{vice versa}). Comparing these plots to the phase maps in Section \ref{sec:tls}, it is apparent that only the tetra- and hemi-tetra- domains ($\xi < 1$) exhibit a dipole response - which is appropriate for our model as $E_{01}$ splittings representative of a TLS only appear in these regions. Localised oxygen atoms ($\xi > 1$) are also expected to not elicit dipole behaviour.

With this information, the domain boundaries and bifurcations on each phase map can now be fully explained. Two variables alter the landscape: dipole and potential. Clusters with tight $z$ confinement (those without tetra-well regions such as $\abs{Z} = 2.25 \; \rm{\AA}$) have four unique regions where a TLS may reside. The dipole direction dominates two of these domains: an A type region when the confining pair is collinear to the dipole, and a B type region when the confining pair is perpendicular. A symmetry bifurcation (at $\abs{X} = \abs{Y}$) separates the dipole domains into four regions which can therefore be clearly identified in terms of dipole moment ($\abs{\wp_x}$ or $\abs{\wp_y}$) and defect type (A or B).

Clusters without as much $z$ confinement (such as $\abs{Z} = 2.75 \; \rm{\AA}$) exhibit tetra-well behaviour: generating two additional regions. Tetra-well domains, as discussed in Subsection \ref{sec:phaseanalysis}, are caused when the confining pair of aluminium atoms start interacting with the defect pair (and hence the oxygen as well). If we consider the A type, $\abs{\wp_y}$ domain in Figure \ref{fig:dipole275}, it is clear that the dominant dipole direction remains constant as $\abs{X}$ separation is in increased and the tetra-well domain is entered. The same $\abs{X},\abs{Y}$ parameters on Figure \ref{fig:dipole225} cross a bifurcation line, changing dipole direction and the model indicates B type defect properties. Increased confinement in $z$ induces a deeper potential well in the $xy$ plane and removes any major landscape changing contributions from the confining atom pair -- effectively reducing the model back to a 1D description. A cluster with a comparatively shallow potential that generates tetra-well domains does so when conditions are advantageous for an A type confinement pair to become a B type defect pair (a dipole direction change is not required for this to occur). A trace like the $\abs{Y} = 2.5 \; \rm{\AA}$ (labelled A) on Figure \ref{fig:xiunbound} (and an associated eigenspectrum response similar to Figure \ref{fig:tetraspectrum}) is an example of this behaviour. If we do not consider the small portion of quad-degenerate defects in this domain; the measurable properties (i.e.\ $E_{01}$ and dipole strength) tetra-well, B type, $\abs{\wp_y}$ TLSs are identical to B type, $\abs{\wp_y}$ TLSs that reside in a hemi-tetra-well domain.

%If however, a configuration lies inside a tetra-well domain after the pair separation on the defect axis has been increased from an A type point, the system cannot change to a B type through an (an)harmonic process and is forced to stay in a regime with the same dipole direction. Figure \ref{fig:dipole275} illustrates this phenomena when $\abs{Y} = 2.7 \; \rm{\AA}$ and $\abs{X}$ is extended from $1.6$ to $2.1 \; \rm{\AA}$.

%float fix position

\section{Qubit Coupling}\label{sec:smax}

To compare our TLS model directly to experiments, we assume that our JJ lies within a phase qubit, although the model applies equally for any device comprised of amorphous junctions. The measurable signal of a TLS in a phase qubit is the resonance of the TLS and qubit splitting energy, $E_{01}$, with the qubit-TLS coupling. For the phase qubit~\cite{Martinis2005}, the qubit-TLS coupling $S_{max}$ is a function of $E_{01}$ and $\wp$~\cite{Kofman2007}, the effective dipole moment due to an electric field applied in the direction of delocalization,
%[\textcolor{red}{Chalmers SET expt}] could be cited here (amorphous junctions)
\begin{equation}
    S_{max}=2\frac{\wp}{w}\sqrt{\frac{e^2}{2C}E_{01}}.
\label{eq:smax}
\end{equation}
Throughout this discussion we assume a junction width $w = 2$ nm and capacitance $C = 850$ fF.

The dipole magnitudes in Figures \ref{fig:dipole275} and \ref{fig:dipole225} are calculated against the electron charge for simplicity, although as we are discussing an oxygen atom, the dipole elements may in fact be larger. Using our Josephson junction DFT models~\cite{DuBois2013} we partition the charge density associated with atoms across the lattice into Bader volumes \cite{Tang2009}. The charge enclosed within each Bader volume is a good approximation to the total electronic charge of an atom. An average value of $1.395\pm0.006 \; e$ is found for oxygen atoms in a junction comprised of AlO$_{0.5}$ at a density of $0.8$ times that of corundum (a common, low temperature and pressure phase of $\mathrm{Al_2O_3}$). We can use this value such that
\begin{equation}
\widetilde{\wp_x} = 1.4e\wp_x
\end{equation}
(for a dipole in the $x$ direction) to gain a better estimate of $S_{max}$.

\begin{figure}[t]
\centering
\includegraphics[width=\columnwidth]{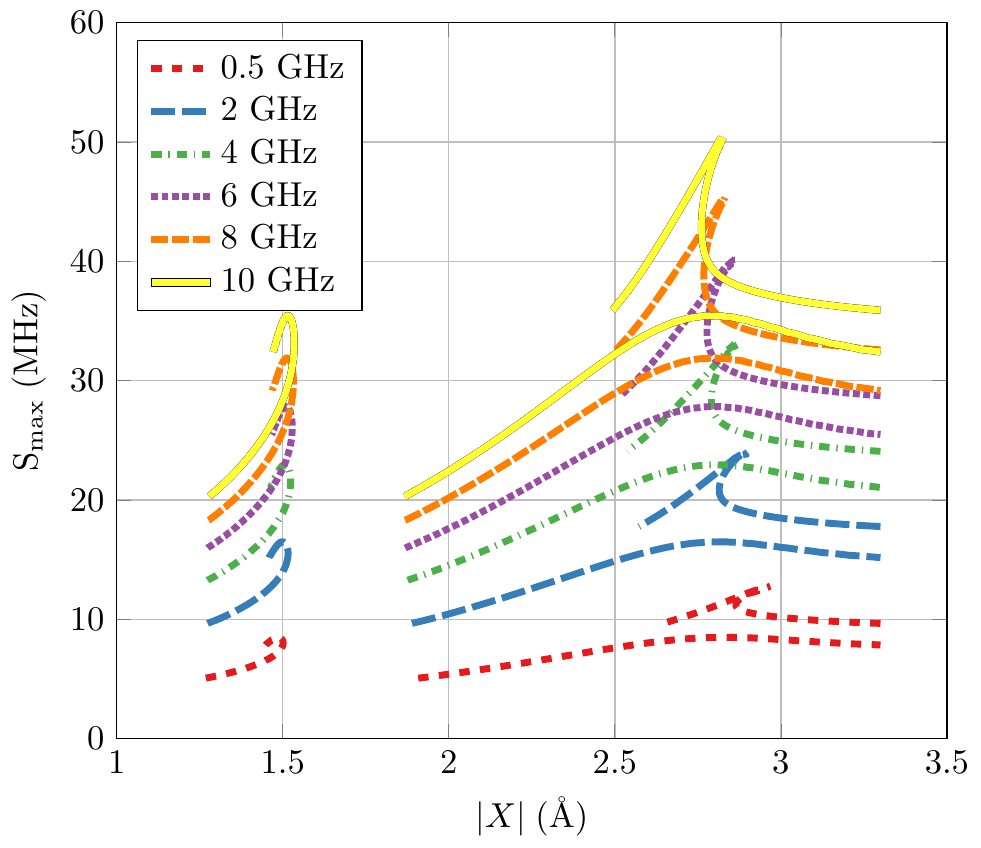}\\
\caption{\label{fig:smax225}(Color online) Coupling strength to a fictitious phase-qubit $S_{max}$ as a function of $\abs{X}$ in the domains where both dipoles $\abs{\wp_{x,y}}$ are dominant (see Eq.\ \ref{eq:smax}) for a set of constant $E_{01}$ splitting frequencies and $\abs{Z} = 2.25 \; \rm{\AA}$.}
\end{figure}

In Figure \ref{fig:smax225} we plot contour lines representing constant values of $E_{01}$ which correspond to the purview of experimentally observed qubit resonant frequencies for constant values of $\abs{Z} = 2.25 \; \mathrm{\AA}$. The $S_{max}$ (Equation \ref{eq:smax}) response to these frequencies is plotted as a function of $\abs{X}$, in which we see maximum coupling strengths which correspond exceptionally well with experimental observations~\cite{Lupascu2009, Shalibo2010, Cole2010}. The most comprehensive of these studies (Ref \onlinecite{Shalibo2010}) measures $S_{max}$ values of $3$--$45$ MHz, assuming a $1 \; \rm{\AA}$ dipole moment.

\begin{figure}[htp]
\centering
\includegraphics[width=\columnwidth]{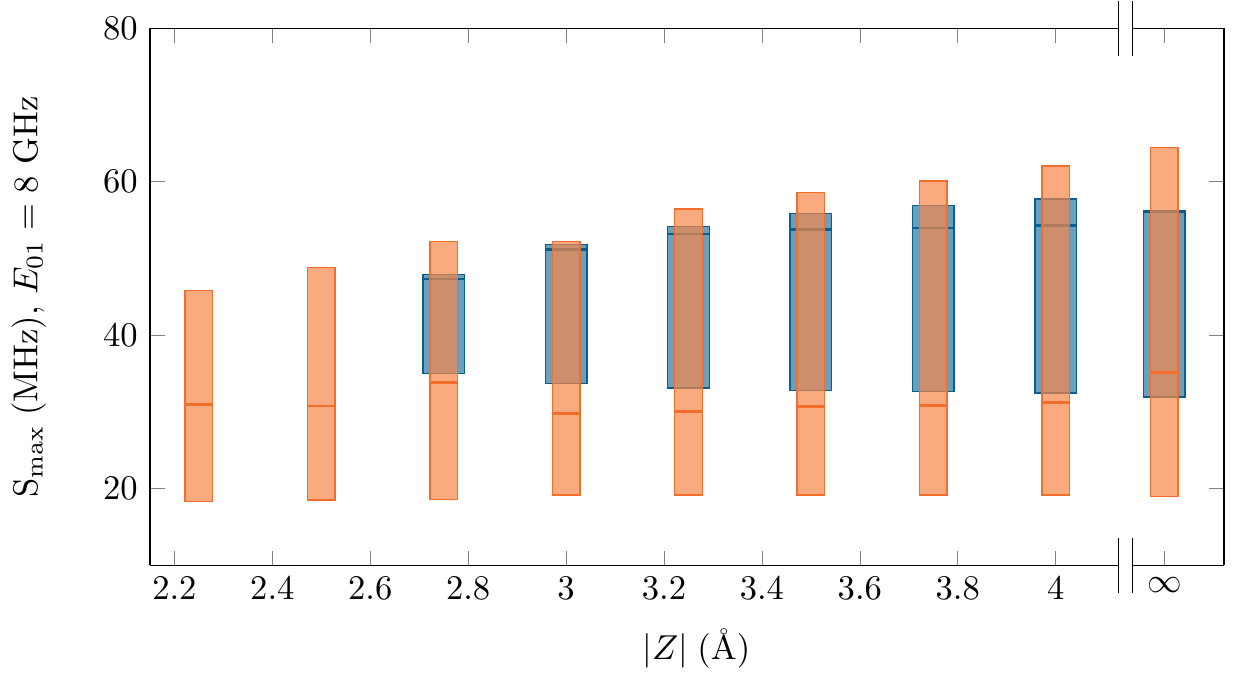}\\
\caption{\label{fig:smaxz}(Color online) $S_{max}$ response range for a constant $E_{01}$ splitting frequency of $8$ GHz as a function of $\abs{Z}$ separation. Each box on the figure represents the minimum to maximum $S_{max}$ values in both $\abs{\wp_{x}}$ and $\abs{\wp_{y}}$ dominant domains over the phase space of $\abs{X}$ distance separations. The box at $\abs{Z} = 2.25 \; \rm{\AA}$ represents all $E_{01} = 8$ GHz values on Figure \ref{fig:smax225} for example. Median values are plotted as solid lines inside each box. Orange (thin) boxes represent quasi-degenerate regions and blue (thick) boxes indicate quad-degenerate regions.}
\end{figure}

Whilst the $S_{max}$ response is neither smooth nor singular over the phase space investigated, the value range is surprisingly small. Figure \ref{fig:smaxz} shows the range of $S_{max}$ couplings for all $E_{01}=8$ GHz configurations calculated with confinements in $\abs{Z}$ from $2.25 \; \rm{\AA}$ to unbound and $\abs{X}$ (hence $\abs{Y}$ as well from symmetry arguments) from $1.2$ to $4 \; \rm{\AA}$. The entire range of $S_{max}$ values is only $60$ MHz wide, which suggests an explanation as to why large couplings (of order $500$ MHz for example) have not been seen experimentally.

\section{Strain Response}\label{sec:strain}

Unusually long coherence times of strongly coupled defects are a key observation of TLS-qubit experiments~\cite{Neeley2008, Lisenfeld2010a}. As our model assumes a charge-neutral defect, coherence time is linked to the dipole element (for charge noise) and the strain response (for phonons). Mechanical deformation of a phase-qubit has recently been observed directly~\cite{Grabovskij2012}, which we can mimic in our 2D model by introducing a series of deformations onto the cluster, and subsequently measure the $E_{01}$ response.

\begin{figure}[t!]
\centering
\includegraphics[width=\columnwidth]{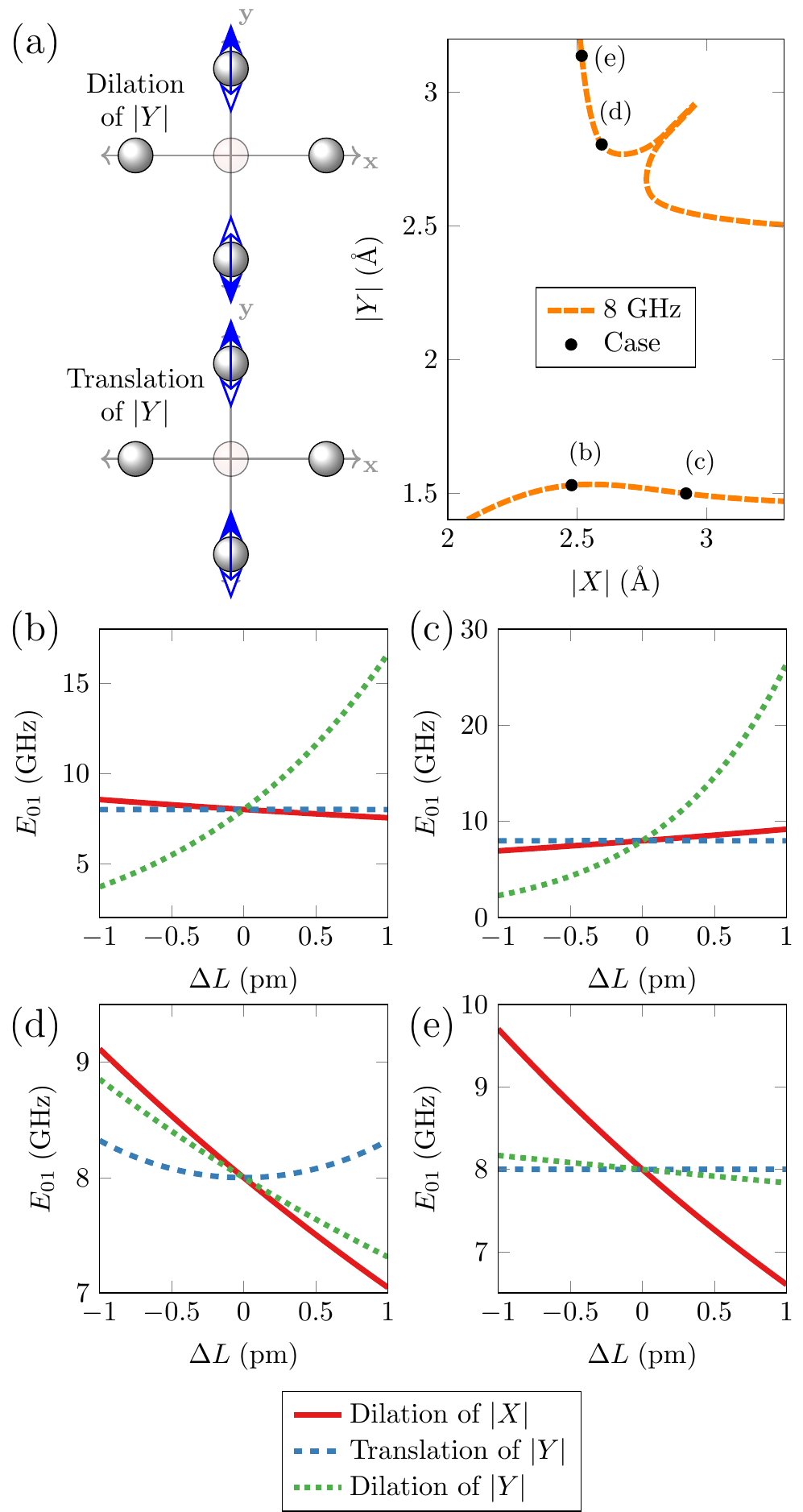}\\
\caption{\label{fig:strain}(Color online) Strain response of four candidate TLS clusters. (a) Left: depictions of two deformation types that show active responses. A dilation: where an $\abs{Y}$ (shown) or $\abs{X}$ pair is stretched from their original position and a translation: where an $\abs{Y}$ (shown) or $\abs{X}$ pair is moved from their original position. Right: locations along the $8$ GHz contour from the $\abs{Z} = 2.25 \; \rm{\AA}$ phase maps. Each case is deformed and translated in both $x$ and $y$ directions by $\pm 1$ pm and plotted as (b), (c), (d) and (e). Note that all four cases lie in the $\abs{\wp_x}$ dominant domain (see Figure \ref{fig:dipole225}). Translation of $\abs{X}$ is not shown on any graph as its' $E_{01}$ response is of $\mathcal{O}(10^4)$ over ($\Delta L$) for all cases.}
\end{figure}

Whilst many deformation types were tested~\cite{DuBois2013}, only two types (depicted in Figure \ref{fig:strain}a) show an active response over a length change ($\Delta L$) of $1$ pm. Four clusters are chosen that lie on the $8$ GHz contour when $\abs{Z} = 2.25 \; \rm{\AA}$, indicated as (b), (c), (d) and (e) in Figure \ref{fig:strain}a. As can be seen in Figure \ref{fig:dipole225}, each of these cases are within a $\abs{\wp_x}$ dominant domain.

The investigation in Ref \onlinecite{DuBois2013} focused on $\abs{X}$ translation due to its large, hyperbolic response -- which is also seen in experiments~\cite{Grabovskij2012}. The cluster configuration considered had a $\abs{Z}$ confinement of $2.5 \; \rm{\AA}$ and the same $\abs{X}$, $\abs{Y}$ separations as published here; thus it is not surprising that $\abs{X}$ translation yields an $E_{01}$ response is of $\mathcal{O}(10^4)$ GHz for $\abs{Z} = 2.25 \; \rm{\AA}$ also. Therefore, translation of $\abs{X}$ has not been included in Figure \ref{fig:strain} or the following discussion.

Clusters (b) and (c) are both identified as A type defects and are relatively insensitive to $\abs{X}$ dilation and $\abs{Y}$ translation. Dilation of $\abs{Y}$ for these defects is a different way of saying ``extending the defect pair separation'' -- which has been described in the above sections. The intensity of the response however is noticeably larger as the confinement distance ($\abs{X}$) is increased.

B type defects, labelled (d) and (e), respond discordantly depending on their configuration. Dilation in $\abs{Y}$, whilst a sizable strain source for A type configurations, initiates no response from the (e) configuration at all. B type defects located at this position in phase space have their defect pair in $\abs{X}$, and point (e) specifically is confined with $\abs{Y} = 3.1 \rm{\AA}$. As discussed in the sections above, this separation distance is close to being practically unbound. Point (d) on the other hand has a tighter separation distance and begins to confine the system. Dilation in $\abs{X}$ responds in the opposite manner effectively: expanding defect separation distance whilst keeping the confinement distance static ultimately confines the wavefunction. The final response that the B type defects respond to is $\abs{Y}$ translation. As point (e) is essentially unbound in $\abs{Y}$ we do not expect any response from a strain of this magnitude. Point (d) on the other hand is interacting with its tighter confinement distance, with the TLS dipole collinear to the $\abs{X}$ direction. Translation of $\abs{Y}$ forces local minima of the potential landscape from points on the $x$ axis to locations off axis, yet still within the minima rings about the aluminium atoms. In other words, the wavefunction is slightly rotated around the defect axis.\\

\section{Conclusions and Outlook}

Even with the extreme idealisation of this model, it allows the prediction of experimentally measured properties of strongly coupled TLSs with atomic positions as the only input parameters and therefore shows as a proof of concept that these defects can arise in AlO$_{x}$ without any alien species present.

Our model shows that even when only considering delocalisation in two dimensions, a range of different behaviour can be seen. The existence of effective two-state systems can come about through different potential shapes, which in turn arise due to various atomic position configurations. To understand these different configurations, we consider both the symmetries of the eigenspectrum and the effective charge dipole of the defect.

We find that two-level systems with equivalent properties to those seen in experiments can be formed from atomic configurations which are entirely consistent with the material properties of amorphous aluminium oxide barriers. Most interestingly we find that the expected qubit-TLS coupling strength and the TLS strain response correspond very well to that observed experimentally.

A complication that occurs with phenomenological models that attempt to describe TLS behaviour is their free parameter count is high enough to describe all facets of the observed experimental parameters without being distinguishable from other, non-complimentary models~\cite{Cole2010}. Whilst this delocalised oxygen model only uses one type of parameter as input (atomic positions), it still requires an $x, y, z$  coordinate set for up to $6$ cage atoms. More sophisticated modelling techniques such as molecular mechanics and density functional methods can be then used to obtain more realistic values for the atomic positions~\cite{DuBois2013}. Microscopic models of this type will guide future fabrication and design of superconducting circuits, leading to lower levels of noise and greater control over their quantum properties.

\begin{acknowledgments}
This research was undertaken with the assistance of resources provided at the NCI National Facility systems at the Australian National University through the National Computational Merit Allocation Scheme supported by the Australian Government.
\end{acknowledgments}

\bibliography{library}

\end{document}